\documentclass[10pt,journal,compsoc]{IEEEtran}
\usepackage{amsmath,amssymb,amsfonts}
\usepackage{graphicx}
\usepackage{algorithm}
\usepackage{algpseudocode}
\usepackage{bm}
\usepackage{color}
\usepackage{amsthm}

\newtheorem{Remark}{Remark}

\newtheorem{Prop}{Proposition}
\newtheorem{Lem}{Lemma}

\def\pbf{{\bf p}}

\def\rbf{{\bf r}}

\def\wbf{{\bf w}}
\def\xbf{{\bf x}}

\def\zbf{{\bf z}}
\def\rbf{{\bf r}}
\def\xbf{{\bf x}}

\def\Zbf{{\bf Z}}

\ifCLASSOPTIONcompsoc
  \usepackage[nocompress]{cite}
\else
  \usepackage{cite}
\fi

\begin{document}

\title{Resource Sharing of a Computing Access\\ Point for Multi-user Mobile Cloud Offloading \\with
            Delay Constraints}

\author{Meng-Hsi Chen,~\IEEEmembership{Student Member,~IEEE,}
        Min Dong,~\IEEEmembership{Senior Member,~IEEE,}
        and~Ben Liang,~\IEEEmembership{Fellow,~IEEE}% <-this % stops a space
\IEEEcompsocitemizethanks{\IEEEcompsocthanksitem Meng-Hsi Chen and
Ben Liang are with the Department of Electrical and Computer
Engineering,
University of Toronto, Toronto, Canada (e-mail: $\{$mchen, liang$\}$@ece.utoronto.ca).\protect\\
Min Dong is with the Department of Electrical, Computer and Software
Engineering, University of Ontario Institute of Technology, Oshawa,
Canada (e-mail: min.dong@uoit.ca). \IEEEcompsocthanksitem This work
has been funded in part by grants from the Natural Sciences and
Engineering Research Council (NSERC) of Canada and in part by the
Ontario Ministry of Research and Innovation under an Early
Researcher Award. \IEEEcompsocthanksitem A preliminary version of
this work has appeared in \cite{chen2016icassp}.
}}% <-this % stops an unwanted space

\IEEEtitleabstractindextext{%
\begin{abstract}
We consider a mobile cloud computing system with multiple users, a
remote cloud server, and a computing access point (CAP). The CAP
serves both as the network access gateway and a computation service
provider to the mobile users. It can either process the received
tasks from mobile users or offload them to the cloud. We jointly
optimize the offloading decisions of all users, together with the
allocation of computation and communication resources, to minimize
the overall cost of energy consumption, computation, and maximum
delay among users. The joint optimization problem is formulated as a
mixed-integer program. We show that the problem can be reformulated
and transformed into a non-convex quadratically constrained
quadratic program, which is NP-hard in general. We then propose an
efficient solution to this problem by semidefinite relaxation and a
novel randomization mapping method. Furthermore, when there is a
strict delay constraint for processing each user's task, we further
propose a three-step algorithm to guarantee the feasibility and
local optimality of the obtained solution. Our numerical results
show that the proposed solutions give nearly optimal performance
under a wide range of parameter settings, and the addition of a CAP
can significantly reduce the cost of multi-user task offloading
compared with conventional mobile cloud computing where only the
remote cloud server is available.
\end{abstract}

\begin{IEEEkeywords}
Mobile cloud computing, computing access point, offloading decision,
resource allocation, energy cost, computation cost, delay cost.
\end{IEEEkeywords}}

\maketitle

\IEEEdisplaynontitleabstractindextext

% \fi
\IEEEpeerreviewmaketitle

\IEEEraisesectionheading{\section{Introduction}\label{sec:introduction}}


\IEEEPARstart{M}{obile} Cloud Computing (MCC) extends the
capabilities of mobile devices to improve user
experience\cite{kumar2013}\cite{fernando2013}\cite{dinh2013}. Mobile users can
offload tasks to the cloud, using abundant cloud resources to help
them gather, store, and process data. However, the interaction
between mobile devices and the cloud introduces some key challenges
in the system design. For example, the decision on whether to offload tasks to the
cloud needs to balance the tradeoff between energy consumption and
computing performance. Furthermore, the communication delay between
mobile users and the cloud needs to be taken into consideration \cite{kumar2013}.

With an aim to reduce the communication delay in task offloading,
 Mobile Edge
Computing (MEC), as defined by the European Telecommunications
Standards Institute (ETSI), is a distributed MCC system where
computing resources are installed locally at or near the base
station of a cellular network
\cite{etsi2016framework}\cite{liang2017}\cite{tran2017}. MEC shares
similarities with micro cloud centers \cite{greenberg2008},
cloudlets\cite{satyanarayanan2009}, cyber-foraging\cite{lewis2015},
and fog computing \cite{bonomi2012}, except that the MEC computing
servers are managed by a mobile service provider, which allows more
direct control and resource management.

Similar to the concept of MEC, in this work, we use the general term
\textit{computing access point (CAP)}, which refers to a wireless
access point or a cellular base station with built-in computation
capability. For example, CAPs may be provided by Internet service
providers as a value-added service. Mobile devices that wish to
offload a task first sends it to the CAP. The CAP may serve its
conventional networking function and forward the task to the remote
cloud server, or directly process the task by itself. The additional
option of computation by the CAP reduces the need for access to the
remote cloud server, and hence can potentially decrease the
communication delay and also the overall energy and computation
cost. However, the availability of CAP adds an extra dimension of
variability for offloading decisions. Each task may be processed
locally at the mobile device, at the CAP, or at the remote cloud
server. Furthermore, both computation and communication resources
need to be considered in different offloading choices. This makes
optimizing the mobile
task offloading decision even more challenging. %In a
%preliminary study \cite{chen2015spawc}, we have demonstrated the
%impact of a CAP on system performance in a \textit{single-user}
%scenario.

In this work, we study the interaction among multiple users, the
CAP, and the cloud.
% between \textit{multiple users} and the
%CAP. We jointly consider both the offloading decision and resource
%allocation among all users. Different from \cite{chen2015spawc}, in
%this
In a multi-user scenario, to offload tasks, we need to allocate
communication and computation resources among competing users. We
jointly consider both the offloading decision and resource
allocation among all users, with an aim to conserve energy and
maintain service quality for all of them. For this joint
optimization problem, an optimal offloading decision must take into
consideration the computation and communication energies,
computation costs, and communication and processing delays at all
local user devices, as well as the resource constraints and
capabilities of the CAP and the remote cloud. The contributions of
this work are summarized as follows:
\begin{itemize}
\item We focus on jointly
optimizing the offloading decisions as well as the computation and
communication resource allocation for multiple mobile users with one
CAP and one remote cloud server. We formulate the joint optimization
problem to minimize a weighted sum of costs of energy, computation,
and the maximum delay among all users. This results in a mixed
integer programming problem. To solve this challenging problem, we
first reformulate and transform the problem into a non-convex
quadratically constrained quadratic program (QCQP) \cite{boyd2004},
which is still NP-hard in general.
 To obtain a solution to this problem, we then propose an efficient heuristic
algorithm, termed \textit{shareCAP}, based on semidefinite
relaxation (SDR) \cite{luo2010} and a novel randomization mapping
method.

\item We further study the scenario where there is a strict processing deadline for each user's task.
With these additional delay constraints, the proposed
\textit{shareCAP} method can no longer be directly applied to find a
solution due to the absence of a feasibility guarantee.
 To solve this more complicated optimization problem, we further propose a three-step
algorithm named \textit{shareCAP-D}, consisting of SDR, adaptive
adjustment, and sequential tuning, to iteratively find a solution.
We show that \textit{shareCAP-D}
guarantees a locally optimal solution.

\item Through numerical study, by comparing with an
optimal offloading policy obtained by exhaustive search, we
demonstrate that the proposed \textit{shareCAP} and
\textit{shareCAP-D} methods give nearly optimal performance under a
wide range of parameter settings. Furthermore, we observe that the
addition of a CAP can significantly reduce the energy and
computational costs of the system, as
compared with the conventional MCC where only the remote cloud
server is available for task offloading.
\end{itemize}

The rest of this paper is organized as follows. Related works are
reviewed in Section \ref{sec_related_work}. In Section
\ref{sec_system_model}, we describe the system model for mobile
cloud computing with a CAP and formulate the optimization problem.
In Section \ref{sec_shareCAP_solution}, we transform our problem to
a QCQP problem and solve it through the SDR approach. In Section
\ref{sec_delay}, we further study the scenario with strict delay
constraints. In Section \ref{sec_sum_delay}, we extend our work to
sum delays optimization. Numerical results are presented in Section
\ref{sec_simulation}, followed by conclusion in Section
\ref{sec_conclusion}.

Notations: We denote by $\mathbf{a}^T$ and $\mathbf{A}^T$ the
transpose of vector $\mathbf{a}$ and matrix $\mathbf{A}$,
respectively. The notation $\mathrm{diag}(\mathbf{a})$ denotes the
diagonal matrix with diagonal elements being elements of vector
$\mathbf{a}$. The trace function of matrix $\mathbf{A}$ is denoted
by $\mathrm{Tr}(\mathbf{A})$. We use $\mathbf{A}(i,j)$ to denote the
$(i,j)\mathrm{th}$ entry of matrix $\mathbf{A}$. We use $\mathbf{A}
\succeq 0$ to indicate that $\mathbf{A}$ is a positive semi-definite
matrix.

%========================================================================
\section{Related Work}\label{sec_related_work} Many existing
works study task offloading from mobile users to the local (or
remote) processor in two-tier cloud systems. For a single mobile
user offloading its entire application to the cloud, the authors of
\cite{kumar2010,zhang2013,wen2012} presented different energy models
to analyze whether or not to offload application to the cloud, and
the tradeoff between energy consumption and computing performance
was studied in \cite{barbarossa2013,munoz2015}. Furthermore, many
studies have considered partitioning an application into multiple tasks. Among
them, MAUI \cite{cuervo2010}, Clonecloud \cite{chun2011}, and Thinkair
\cite{kosta2012} are systems proposed to enable a mobile device to
offload tasks to the cloud. These works focus on the implementation
of offloading mechanisms from the mobile device to the cloud, and
the discussion on optimizing the offloading decisions was limited.
%The design of offloading decisions was considered in \cite{zhang2012} to \cite{Truong-Huu2014}.
In \cite{zhang2012} and \cite{zhang2013infocom}, heuristic
offloading policies were proposed for a mobile user with sequential
tasks.
 In \cite{mahmoodi2016,kao2015,wu2016}, the problem of cloud offloading for a mobile user
with dependent tasks was studied. In \cite{zhang2015}, offloading a
mobile user's tasks in an intermittently connected cloud system was
considered.
The impact of mobility was considered in \cite{Truong-Huu2014},
where the authors proposed an opportunistic offloading algorithm.
All of the studies above focus on a single mobile user.

Task offloading by multiple mobile users have been considered in
\cite{hoang2012,Hoang2014,kaewpuang2013,ren2013,sardellitti2015,lyu2017,chen2014,meskar2017,chen2015efficient},
where each user has a single application or task to be offloaded to
the cloud in its entirety. In
\cite{hoang2012,Hoang2014,kaewpuang2013}, the authors considered
optimizing offloading decisions, aiming to maximize the revenue of
the mobile cloud service providers under a fixed resource usage per
user. The cooperation among selfish service providers to improve the
revenue was further studied in \cite{kaewpuang2013}.
 The authors
of \cite{ren2013,sardellitti2015} studied the allocation of radio
and computation resources in the scenario where all tasks are always
offloaded. The joint optimization of offloading decision and
communication and computation resources for system utility
maximization was considered in \cite{lyu2017}, where where the
number of tasks that can be offloaded is limited by the transmission
bandwidth; a heuristic algorithm  was proposed to obtain the
resource allocation and offloading decision sequentially.
 Game theoretic
approaches were adopted in
\cite{chen2014,meskar2017,chen2015efficient} to study decentralized
decision control in systems where offloading decisions are made by
mobile users as selfish players. However, these game theoretic works
focus on the offloading decisions for each user without considering
the allocation of communication and computation resources.
Furthermore, a multi-user scenario where each user has multiple
independent tasks was considered in \cite{chen2016icc}, where the
offloading decision algorithm were proposed by minimizing the
weighted cost of energy consumption and worst-case offloading delay.
The authors of \cite{viswanathan2016} considered a mobile device
cloud, which is composed purely of proximal mobile devices, and a
task scheduling mechanism was proposed for concurrent application
management. Coordination of local mobile devices forming a mobile
cloud has been studied in \cite{Habak2015}. All of the studies above
focus on a two-tier cloud network consisting of only mobile users
and another tier of local or remote processors.

The three-tier network consisting of mobile users, a local computing
node (e.g., cloudlet or CAP), and a remote cloud server has been
studied in
\cite{Rahimi2012,Rahimi2013,Song2014,cardellini2016,chen2017infocom}.
Without considering resource allocation, centralized heuristic
algorithms for offloading decisions were proposed in
\cite{Rahimi2012,Rahimi2013,Song2014}, while a game theoretic
approach was considered to distributedly obtain the offloading
decision in \cite{cardellini2016}. Despite these works, the joint
optimization of the offloading decision and the allocation of
computation and communication resources for a general three-tier
system has not been investigated before. The joint optimization
problem is much more complicated to solve, because the offloading
decision and resource allocation are inter-dependent.

In our recent work \cite{chen2017infocom}, a multi-user scenario
where each user has multiple independent tasks was considered for
joint optimization of offloading and allocation of communication and
computation resources. The differences of this work and
\cite{chen2017infocom} are as follows: 1) The problem structures are
different, leading to different problem formulations and solution
approaches; 2) For the single-task per user case studied in this
paper, we propose a low-complexity algorithm that is shown to
achieve nearly optimal performance. This combined advantage in both
the complexity and performance cannot be achieved by the algorithm
proposed in \cite{chen2017infocom}; 3) In this work, we further
study the scenario where a strict processing deadline is imposed on
each user's task, which cannot be addressed by the solution approach
proposed in \cite{chen2017infocom}.

%========================================================================
%\vspace{-0.3cm}
\section{System Model and Problem Formulation}\label{sec_system_model}
In this section, we first introduce the model of mobile cloud
computing with a CAP, detailing the costs of processing locally, at
the CAP, and at the cloud. We then explain the joint offloading
decision and resource allocation optimization problem to minimize a
weighted sum cost.

\subsection{System Model}
\subsubsection{Mobile Cloud with CAP}
\begin{figure}[t]
%\vspace{-0.1cm}
\centering
\includegraphics [scale =0.31]{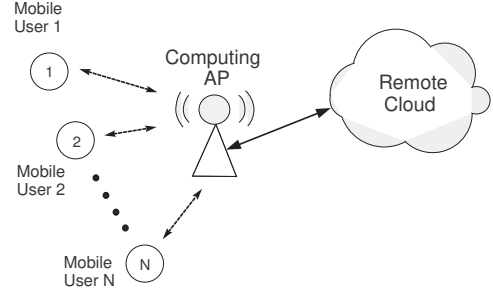}
\vspace{-0.2cm} \caption{System model} \label{fig:system_model}
\vspace{-0.3cm}
\end{figure}

Consider a cloud access network with $N$ mobile users, one CAP, and
one remote cloud server, as shown in Fig.~\ref{fig:system_model}.
The CAP is a wireless access point (or a cellular base station) with
built-in computation capability that may be provided by Internet
service providers as a value-added service. Instead of just serving
as a relay to always forward received tasks from users to the cloud,
the CAP also has the capability to process user tasks subject to its
resource constraint.
%We assume all mobile users are within the
%service coverage of the CAP during the whole transmission and
%processing period, and
 We denote
the set of all users by $\mathcal{N}=\{1,...,N\}$. Each mobile user
has one task to be either processed locally or offloaded to the CAP,
and the CAP determines whether to process each received task by
itself or further offload it to the cloud for processing. Since
there are multiple tasks offloaded to the CAP and some of them are
processed by the CAP, we need to further allocate the communication
and computation resources available at the CAP.

We assume that all tasks are available at time zero. This is similar
to many existing studies
\cite{zhang2013,wen2012,barbarossa2013,munoz2015,sardellitti2015,chen2014,meskar2017,chen2015efficient,lyu2017}.
If the tasks arrive dynamically in time, we may apply our model and
the proposed solution in a quasi-static manner, where the system
processes the tasks in batches that are collected over time
intervals \cite{Shmoys1995}.

\subsubsection{Offloading Decision}
Denote the
offloading decisions for user $i$ by $x^l_{i}, x^a_{i}, x^c_{i} \in
\{0,1\}$, indicating whether user $i$'s task is processed locally,
at the CAP, or at the cloud, respectively.  The offloading decisions
are constrained by
\begin{equation}\label{a}
x^l_{i} + x^a_{i} + x^c_{i} = 1,\ i\in\mathcal{N}.
\end{equation}
Notice that only one of $x^l_{i}$, $x^a_{i}$, and $x^c_{i}$ for user
$i$ can be $1$.
%\vspace{-0.3cm}
\subsubsection{Cost of Local Processing}
%\vspace{-0.1cm}
The input data size, output data size, and processing cycles of user
$i$'s task are denoted by $D_{{\textrm{in}}}(i)$,
$D_{{\textrm{out}}}(i)$, and $Y(i)$\footnote{The processing cycles
of user $i$'s task depends on the input data size and the
application type. For simplicity of illustration, we initially
assume that a task requires the same value of $Y(i)$ on different
CPUs so that its processing time is a function of the CPU's clock
speed only. We will explain later how the proposed solution can be
trivially extended to the general case.}, respectively. Similar to
\cite{wen2012,barbarossa2013,munoz2015,sardellitti2015,chen2014,meskar2017,chen2015efficient,lyu2017},
 we assume that these quantities are known,
 which may be achieved by applying
a program profiler \cite{cuervo2010, chun2011,kosta2012}. We assume
that the additional instructions required for remote processing can
be downloaded directly by the CAP or the cloud via their access to a
high-capacity wired network.
 When the task is
processed locally, the processing energy is denoted by $E^l_{i}$ and
the processing time is denoted by $T^l_{i}$.
\subsubsection{Cost of CAP Processing}
For user $i$'s task being offloaded to the CAP, we denote the energy
consumed by wireless transmission (to the CAP) and reception (from
the CAP) at user $i$ by $E^t_{i}$ and $E^r_{i}$, respectively. We
further denote the uplink and downlink transmission times between
user $i$ and the CAP by
$T^t_{i}=D_{{\textrm{in}}}(i)/(\eta^u_{i}c^u_{i})$ and
$T^r_{i}=D_{{\textrm{out}}}(i)/(\eta^d_{i}c^d_{i})$, respectively,
where $c^u_{i}$ and $c^d_{i}$ are uplink bandwidth and downlink
bandwidth allocated to user $i$, and $\eta_i^u$ and $\eta_i^d$ are
the spectral efficiency of uplink and downlink transmission between
user $i$ and the CAP, respectively\footnote{The spectral efficiency
can be approximated by $\log(1+\mathrm{SNR})$ where $\mathrm{SNR}$
is the link quality between user $i$ and the CAP.}. Furthermore,
$c^u_{i}$ and $c^d_{i}$ are limited by the uplink bandwidth
$C_{\textrm{UL}}$ and downlink bandwidth $C_{\textrm{DL}}$ as
follows
\begin{align}\label{b}
\sum_{i=1}^Nc^u_{i}\leq C_{\mathrm{UL}},
\end{align}
and
\begin{align}\label{c}
\sum_{i=1}^Nc^d_{i}\leq C_{\mathrm{DL}}.
\end{align}
Since some uplink and downlink transmissions may overlap with each
other, there is also a total bandwidth constraint
\begin{align}\label{cap_total}
\sum_{i=1}^N(c^u_{i}+c^d_{i})\leq
C_{\mathrm{Total}}.
\end{align}
If this task is processed by the CAP, denote its processing time by
$T^a_{i}=Y(i)/f^a_i$, where $f^a_i$ is the assigned processing rate,
which is limited by the total processing rate $f_A$ at the CAP as
\begin{align}\label{d}
\sum_{i=1}^Nf^a_{i}\leq f_{A}.
\end{align}
The usage cost associated with the CAP processing user $i$'s task is
denoted by $C_i^a$. The usage cost may depend on the data size and
processing cycles of a task, as well as the hardware and energy cost
to maintain the CAP. Detailed modeling of the usage cost is outside
the scope of this work. Here we simply assume that $C^a_{i}$ is
given for all $i$.

\subsubsection{Cost of Cloud Processing}

\begin{table}[t]
%\vspace{-0.1cm}
\caption{Notation and corresponding description.} \centering
%\vspace{-0.2cm}
\normalsize
\begin{tabular}{|r| l|}
\hline %\vspace{0.1cm}
\textbf{Notation}& \textbf{Description} \\
\hline %\vspace{0.2cm}
$D_{{\textrm{in}}}(i)$,\ $D_{{\textrm{out}}}(i)$ & input data size and output data size of \\
         & user $i$'s task\\
$Y(i)$ & processing cycles of user $i$'s task\\
$E^l_{i}$ & local processing energy of user $i$'s task\\
$E^t_{i}$,\ $E^r_{i}$ & uplink transmitting energy and \\
         &  downlink  receiving energy of user $i$'s  \\
         & task to and from the CAP\\
$T^l_{i}$, $T^a_{i}$, $T^c_{i}$ & local processing time, CAP processing\\
         & time, and cloud processing time of \\
         & user $i$'s  task\\
$T^t_{i}$,\ $T^r_{i}$ & uplink transmission time and \\
         &  downlink transmission time of \\
         &  user $i$'s task between the mobile user \\
         & and the CAP\\
$T^{ac}_{i}$ & transmission time of user $i$'s task \\
         &  between the CAP and the cloud\\
$C_{\textrm{UL}}$,\ $C_{\textrm{DL}}$ & uplink bandwidth and downlink \\
         &  bandwidth for transmission between \\
         &  mobile users and the CAP\\
$C_{\textrm{Total}}$ & total transmission bandwidth between \\
         &  mobile users and the CAP\\
$c^u_{i}$,\ $c^d_{i}$ & uplink bandwidth and downlink\\
         &  bandwidth assigned to user $i$\\
$\eta_i^u$,\ $\eta_i^d$ & spectral efficiency of uplink and \\
         &  downlink transmission between user $i$ \\
        & and the CAP\\
$C^a_{i}$,\ $C^c_{i}$ & CAP usage cost and cloud usage cost\\
        & of user $i$'s task\\
$r^{ac}$ & transmission rate for each user \\
        & between the CAP and the cloud\\
$f^{c}$ & cloud processing rate for each user\\
$f_A$ &  total CAP processing rate\\
$f^a_i$ & CAP processing rate assigned to user $i$\\
$\alpha$ & weight of the CAP usage cost\\
$\beta$ & weight of the cloud usage cost\\
$\rho_i$ & weight of the energy consumption of \\
         &user $i$'s task\\
 \hline
\end{tabular}
\label{table_notation}
%\label{} % is used to refer this table in the text
%\vspace{-0.5cm}
\end{table}

If a task is further offloaded to the cloud from the CAP, besides
all the costs mentioned above (except $T^a_{i}$ and $C^a_{i}$
related to the task processing cost by the CAP), there is an
additional transmission time between the CAP and the cloud, denoted
by $T^{ac}_{i}=(D_{{\textrm{in}}}(i)+D_{{\textrm{out}}}(i))/r^{ac}$,
where
 $r^{ac}$ is the transmission
rate between the CAP and the cloud. Also, the cloud processing time
is denoted by $T^{c}_{i}=Y(i)/f^{c}$, where $f^c$ is the cloud
processing rate \textit{for each user}. The rate $r^{ac}$ is assumed
to be a pre-determined value regardless of the number of users. This
is because the CAP-cloud link is likely to be a high-capacity wired
connection as compared
 with the limited wireless links between the mobile users
and the CAP, thus there is no need to consider bandwidth sharing
among the users. Similarly, $f^c$ is also assumed to be a
pre-determined value because of the high computational capacity and
dedicated service of the remote cloud server. Thus, $T^{ac}_{i}$ and
$T^{c}_{i}$ only depend on task $i$ itself. Finally, the cloud usage
cost of processing user $i$'s task at the cloud is denoted by
$C^c_{i}$.

The above notations are summarized in Table \ref{table_notation}.

%\vspace{-0.2cm}
\subsection{Problem Formulation}
%\vspace{-0.2cm}
Our goal is to reduce the mobile users' energy consumption and
maintain the service quality to their tasks. To do so, we define the
total system cost as the weighted sum of total energy consumption,
the costs to offload and process all tasks, and the corresponding
maximum transmission and processing delays among all users. We aim
to minimize the total system cost by jointly optimizing the task
offloading decision vector $\mathbf{x}_i = [x^l_i, x^a_i, x^c_i]^T$
and the communication and CAP processing resource allocation vector
$\textbf{r}_i=[c^u_{i},\ c^d_{i},\ f^a_{i}]^T$.

For user $i$'s task being offloaded to the CAP, we define $E^A_i$ as
the weighted transmission energy and processing cost. Similarly, we
define $E_i^C$ as the weighted transmission energy and processing
cost if the task is offloaded to the cloud. They are given by
\begin{align}
E^A_i=(E^t_{i}+E^r_{i}+\alpha C^a_{i}),\allowdisplaybreaks\nonumber
\end{align}
and
\begin{align}
E^C_i=(E^t_{i}+E^r_{i}+\beta C^c_{i}),\allowdisplaybreaks\nonumber
\end{align}
where $\alpha$ and $\beta$ are the relative weights between the
transmission energy and the processing cost in $E_i^A$ and $E_i^C$,
respectively. The local processing delay at user $i$, denoted by
$T^L_{i}$, is given by
\begin{align}
T^L_{i}= T^l_{i}x^l_{i}.\allowdisplaybreaks\nonumber
\end{align}
Also, define $T^A_{i}$ and $T^C_{i}$ as the transmission and
processing delay at the CAP and the cloud, respectively. We have
\begin{align}
T^A_{i}=\left(\frac{D_{{\textrm{in}}}(i)}{\eta^u_ic^u_{i}}+\frac{D_{{\textrm{out}}}(i)}{\eta^d_ic^d_{i}}+\frac{Y(i)}{f^a_i}\right)x^a_{i},\allowdisplaybreaks\nonumber
\end{align}
and
\begin{align}
T^C_{i}=\left(\frac{D_{{\textrm{in}}}(i)}{\eta^u_ic^u_{i}}+\frac{D_{{\textrm{out}}}(i)}{\eta^d_ic^d_{i}}+T^{ac}_i+T^{c}_i\right)x^c_{i}.\allowdisplaybreaks\nonumber
\end{align}
The values of $T^L_{i}$, $T^A_{i}$, and $T^C_{i}$ depend on the
offloading decisions $\mathbf{x}_i$ and the communication and CAP
processing resource allocation $\textbf{r}_i$. The joint
optimization of offloading and resource allocation is formulated as
follows
\begin{align}
\min\limits_{\{ \mathbf{x}_i \}, \{ \mathbf{r}_i \} }
&\quad \bigg[\sum_{i=1}^N\rho_i(E^l_{i}x^l_{i}+E^A_ix^a_{i}+E^C_ix^c_{i})\nonumber\\
&\quad +\max_{i}\{T^L_{i}+T^A_{i}+T^C_{i}\}\bigg]\label{e}\allowdisplaybreaks\\
\text{s.t.} &\quad \eqref{a},\eqref{b},\eqref{c},\eqref{cap_total},\eqref{d},\nonumber\\
&\quad c^u_{i}, c^d_{i}, f^a_{i} \geq 0,\ i\in\mathcal{N},\label{f}\allowdisplaybreaks\\
&\quad x^l_{i}, x^a_{i}, x^c_{i} \in \{0,1\},\
i\in\mathcal{N},\label{g}
\end{align}
where $\rho_i$ is the weight on energy consumption relative to the
delay. The proposed optimization problem \eqref{e} can be solved by any controller in this network after
collecting all required information. In practice, the controller
could be the CAP. That is, each user provides its information
to the CAP, and the CAP will broadcast the obtained offloading
decisions (and the corresponding resource allocations) to all users
by solving problem \eqref{e}.

Notice that in problem \eqref{e}, the cost of delay is considered in
the total system cost objective. We put different emphasis on delay
by adjusting $\rho_i$. Note that, since processing delay is in the
objective instead of as a constraint in problem \eqref{e}, any
offloading decision and resource allocation are feasible. However,
in practice, there are applications that require strict processing
deadlines, and some offloading decisions may not satisfy the strict
delay constraint for a task. This scenario will be further discussed
in Section \ref{sec_delay}.

%========================================================================
%\vspace{-0.3cm}
\section{ShareCAP Offloading Solution}\label{sec_shareCAP_solution}
For the scenario without any delay constraint, we show in this
section that optimization problem \eqref{e} has an equivalent QCQP
formulation that is NP-hard in general. We then present our proposed
solution through the SDR and randomization mapping approach.

\subsection{Overview of the Proposed Solution}

Given some offloading decisions $\textbf{x}_i$, problem \eqref{e}
concerns only the resource allocation vector $\mathbf{r}_i$ as
\begin{align}
\min\limits_{\{\textbf{r}_i\}}
&\quad \bigg(E+\max_{i}\{T^L_{i}+T^A_{i}+T^C_{i}\}\bigg)\label{h}\\
\text{s.t.} &\quad \eqref{b},\eqref{c},\eqref{cap_total},\eqref{d},
\mathrm{and}\ \eqref{f},\nonumber
\end{align}
where
\begin{align*}
E\triangleq\sum_{i=1}^N\rho_i(E^l_{i}x^l_{i}+E^A_ix^a_{i}+E^C_ix^c_{i}).
\end{align*}
Note that $E$ only depends on $\textbf{x}_i$, and thus can be
treated as a constant. The resource allocation problem \eqref{h} is
convex. It can be solved optimally using standard convex
optimization approaches such as the interior-point method. Since
there are a finite number of offloading decisions, a globally
optimal solution for problem \eqref{e} can be obtained by exhaustive
search among $3^N$ possible offloading decisions. However, the
complexity grows exponentially with the number of users and thus
impractical.

In order to find an efficient solution to problem \eqref{e}, we
first transform it into a separable QCQP with a linear objective,
and then propose a separable SDR approach and a novel randomization
mapping method to recover the binary offloading decisions. Once we
obtain the binary offloading decisions, we can easily solve problem
\eqref{h} to find the corresponding optimal resource allocation. We
name our method the
\textit{shareCAP} offloading and resource allocation solution. %\vspace{-0.2cm}
\subsection{QCQP Reformulation and Semidefinite Relaxation}\label{sec_SDR_no_delay}
We first replace the integer constraint \eqref{g} by
\begin{align} \label{xy_new}
x^s_{i}(x^s_{i}-1)=0,\ i\in\mathcal{N},
\end{align}
for $s \in \{l,a,c\}$. Then, we move the delay term from the
objective to the constraints by introducing additional auxiliary
variable $t$. Optimization problem \eqref{e} is equivalent to the
following problem
\begin{align}\hspace*{-0.3cm}
\min\limits_{\{ \mathbf{x}_i \}, \{ \mathbf{r}_i \},\ t}
&\quad \hspace*{-0.15cm}\sum_{i=1}^N\rho_i(E^l_{i}x^l_{i}+E^A_ix^a_{i}+E^C_ix^c_{i})+t\label{problem_new}\allowdisplaybreaks\\
\text{s.t.}
&\quad \hspace*{-0.15cm}T^l_{i}x^l_{i}+\left(\frac{D_{{\textrm{in}}}(i)}{\eta^u_ic^u_{i}}+\frac{D_{{\textrm{out}}}(i)}{\eta^d_ic^d_{i}}+\frac{Y(i)}{f^a_i}\right)x^a_{i}\nonumber\allowdisplaybreaks\\
&\quad \hspace*{-0.25cm}+\hspace*{-0.05cm}\left(\hspace*{-0.05cm}\frac{D_{{\textrm{in}}}(i)}{\eta^u_ic^u_{i}}\hspace*{-0.05cm}+\hspace*{-0.1cm}\frac{D_{{\textrm{out}}}(i)}{\eta^d_ic^d_{i}}\hspace*{-0.05cm}+\hspace*{-0.05cm}T^{ac}_i\hspace*{-0.1cm}+\hspace*{-0.05cm}T^{c}_i\right)x^c_{i}\hspace*{-0.1cm}\leq \hspace*{-0.05cm}t,\  \hspace*{-0.05cm}i\hspace*{-0.05cm}\in\hspace*{-0.05cm}\mathcal{N},\hspace*{-0.05cm}\label{problem_new.a}\allowdisplaybreaks\\
&\quad
\hspace*{-0.15cm}\eqref{a},\eqref{b},\eqref{c},\eqref{cap_total},\eqref{d},\eqref{f},\
\mathrm{and}\ \eqref{xy_new}.\nonumber
\end{align}
We now show that the optimization problem \eqref{problem_new} can be
transformed into a separable QCQP problem by the following steps.

First, we introduce additional auxiliary variables
$\textbf{d}_i\triangleq(D^u_{i},D^d_{i},D^a_{i})$. Constraint
\eqref{problem_new.a} can be equivalently replaced by the following
four constraints
\begin{align}\label{constraint_delay}
&T^l_{i}x^l_{i}+D^u_{i}+D^d_{i}+D^a_{i}+(T^{ac}_i+T^{c}_i)x^c_{i}\leq
t,\ i\in\mathcal{N},
\end{align}
\begin{align}\label{constraint_D^u}
\frac{D_{{\textrm{in}}}(i)(x^a_{i}+x^c_{i})}{\eta^u_ic^u_{i}}\leq
D^u_{i},\ i\in\mathcal{N},
\end{align}
\begin{align}\label{constraint_D^d}
\frac{D_{{\textrm{out}}}(i)(x^a_{i}+x^c_{i})}{\eta^d_ic^d_{i}}\leq
D^d_{i},\ i\in\mathcal{N},
\end{align}
and
\begin{align}\label{constraint_D^a}
\frac{Y(i)x^a_{i}}{f^a_i}\leq D^a_{i},\ i\in\mathcal{N},
\end{align}
where constraint \eqref{constraint_delay} is the overall delay
constraint, constraints \eqref{constraint_D^u} to
\eqref{constraint_D^a} correspond to the uplink transmission time,
the downlink transmission time, and the CAP processing time,
respectively.

Next, we vectorize the variables and parameters in
\eqref{problem_new}. Define
\begin{align}
\mathbf{w}_0\triangleq[t,\ \mathbf{0}_{1\times 8}]^T,
\end{align}
and
\begin{align}\label{def_w}
 \mathbf{w}_i\triangleq[x^l_{i}, x^a_{i},
x^c_{i},c^u_i,D^u_{i}, c^d_i,D^d_{i},f^a_i,D^a_{i}]^T,\
i\in\mathcal{N},
\end{align}
which is the decision vector for user $i$ containing all decision
variables.
Then, we can rewrite the objective in \eqref{problem_new} as
\begin{align}
\sum_{i=0}^N\mathbf{b}_i^T\mathbf{w}_i,\label{problem_QCQP.obj}
\end{align}
where
\begin{align}
\mathbf{b}_{0}\triangleq[1, \mathbf{0}_{1\times
8}]^T,\nonumber\allowdisplaybreaks
\end{align}
and
\begin{align}
\mathbf{b}_{i}\triangleq\rho_i[E^l_i, E^A_{i}, E^C_{i},
\mathbf{0}_{1\times 6}]^T,\ \mathrm{for}\ i \neq
0.\nonumber\allowdisplaybreaks
\end{align}
In the following, we present each constraint in problem
\eqref{problem_new} in a corresponding matrix form. For the overall
delay constraint \eqref{constraint_delay}, it can be rewritten as
\begin{align}
\sum_{k=0}^N({\mathbf{b}^c_{ik}})^T\mathbf{w}_k\leq 0,\
i\in\mathcal{N},\label{problem_QCQP.a}
\end{align}
where
\begin{align}
&\mathbf{b}^c_{i0}\triangleq[-1, \mathbf{0}_{1\times 8}]^T,\nonumber\allowdisplaybreaks\\
&\mathbf{b}^c_{ii}\triangleq[T^l_{i}, 0, (T^{ac}_i+T^c_{i}), 0, 1, 0, 1, 0, 1]^T,\nonumber\allowdisplaybreaks\\
&\mathbf{b}^c_{ik}\triangleq0,\ \mathrm{for}\ k\neq
0,i.\nonumber\allowdisplaybreaks
\end{align}
The matrix forms of constraints \eqref{constraint_D^u} -
\eqref{constraint_D^a} are
\begin{align}
\mathbf{w}_i^T\mathbf{A}^\mu_{i}\mathbf{w}_i+({\mathbf{b}^\mu_{i}})^T\mathbf{w}_i\leq
0,\ \mu\in\{u,d,a\},\
i\in\mathcal{N},\label{problem_QCQP.b}\allowdisplaybreaks
\end{align}
where
\begin{align}
&\mathbf{A}^{u'}_i\triangleq -0.5
\begin{bmatrix}
0&\eta^{u}_i\\
\eta^{u}_i&0\\
\end{bmatrix},\ \ \nonumber\allowdisplaybreaks
\mathbf{A}^{d'}_i\triangleq -0.5
\begin{bmatrix}
0&\eta^{d}_i\\
\eta^{d}_i&0\\
\end{bmatrix},\nonumber\allowdisplaybreaks\\
&\mathbf{A}^u_i\triangleq
\begin{bmatrix}
\mathbf{0}_{3\times 3}&\mathbf{0}_{3\times 2}&\mathbf{0}_{3\times 4}\\
\mathbf{0}_{2\times 3}&\mathbf{A}^{u'}_i&\mathbf{0}_{2\times 4}\\
\mathbf{0}_{4\times 3}&\mathbf{0}_{4\times 2}&\mathbf{0}_{4\times 4}\\
\end{bmatrix},\nonumber\allowdisplaybreaks\\
&\mathbf{A}^d_i\triangleq
\begin{bmatrix}
\mathbf{0}_{5\times 5}&\mathbf{0}_{5\times 2}&\mathbf{0}_{5\times 2}\\
\mathbf{0}_{2\times 5}&\mathbf{A}^{d'}_i&\mathbf{0}_{2\times 2}\\
\mathbf{0}_{2\times 5}&\mathbf{0}_{2\times 2}&\mathbf{0}_{2\times 2}\\
\end{bmatrix},\nonumber\allowdisplaybreaks\\
&\mathbf{A}^{a'}_i\triangleq -0.5
\begin{bmatrix}
0&1\\
1&0\\
\end{bmatrix},\ \ \nonumber\allowdisplaybreaks
\mathbf{A}^a_i\triangleq
\begin{bmatrix}
\mathbf{0}_{7\times 7}&\mathbf{0}_{7\times 2}\\
\mathbf{0}_{2\times 7}&\mathbf{A}^{a'}_i\\
\end{bmatrix},\nonumber\allowdisplaybreaks\\
&\mathbf{b}^u_i\triangleq[0,D_{\textrm{in}}(i),D_{\textrm{in}}(i), \mathbf{0}_{1\times 6}]^T,\nonumber\\
&\mathbf{b}^d_i\triangleq[0,D_{\textrm{out}}(i),D_{\textrm{out}}(i), \mathbf{0}_{1\times 6}]^T,\nonumber\\
&\mathbf{b}^a_i\triangleq[0,Y(i),0, \mathbf{0}_{1\times
6}]^T.\nonumber\allowdisplaybreaks
\end{align}
We then replace the offloading placement constraint \eqref{a} with
\begin{align}
({\mathbf{b}^P_{i}})^T\mathbf{w}_i= 1,\
i\in\mathcal{N},\label{problem_QCQP.c}\allowdisplaybreaks
\end{align}
where $ \mathbf{b}^P_i\triangleq[1,1,1, \mathbf{0}_{1\times 6}]^T. $
For uplink and downlink bandwidth resource constraints \eqref{b} and
\eqref{c}, we rewrite them as
\begin{align}
\sum_{i=1}^N({\mathbf{b}^U_{i}})^T\mathbf{w}_i\leq
C_{\mathrm{UL}},\label{problem_QCQP.d}\allowdisplaybreaks
\end{align}
and
\begin{align}
\sum_{i=1}^N({\mathbf{b}^D_{i}})^T\mathbf{w}_i\leq
C_{\mathrm{DL}},\label{problem_QCQP.e}\allowdisplaybreaks
\end{align}
where
\begin{align}
&\mathbf{b}^{U}_i\triangleq[0,0,0,1, \mathbf{0}_{1\times
5}]^T,\ \ \mathbf{b}^{D}_i\triangleq[\mathbf{0}_{1\times
5},1,0,0,0]^T.\nonumber
\end{align}
%$\mathbf{b}^{U}_i\triangleq[0,0,0,\gamma^{u}_i, \mathbf{0}_{1\times 5}]^T$ and
%$\mathbf{b}^{D}_i\triangleq[\mathbf{0}_{1\times 5},\gamma^{d}_i,0,0,0]^T$, respectively.
Similarly, the total bandwidth constraint \eqref{cap_total} is as
follows
\begin{align}
\sum_{i=1}^N({\mathbf{b}^{S}_{i}})^T\mathbf{w}_i\leq
C_{\mathrm{Total}},\label{problem_QCQP.total}\allowdisplaybreaks
\end{align}
where
$\mathbf{b}^{S}_i\triangleq[0,0,0,1,0,1,0,0,0]^T.$
The constraint \eqref{d} on the CAP processing resource allocation
can be rewritten as
\begin{align}
\sum_{i=1}^N({\mathbf{b}^A_{i}})^T\mathbf{w}_i\leq
f_A,\label{problem_QCQP.f}\allowdisplaybreaks
\end{align}
where $ \mathbf{b}^{A}_i\triangleq[\mathbf{0}_{1\times 7},1,0]^T. $
Constraint \eqref{f}, which ensures that all variables are
nonnegative, is replaced by
\begin{align}
\mathbf{w}_i \succeq 0,\
i\in\mathcal{N}\cup\{0\}.\label{problem_QCQP.h}\allowdisplaybreaks
\end{align}
Finally, we rewrite integer constraint \eqref{xy_new} as
\begin{align}
\mathbf{w}_i^T\textrm{diag}(\mathbf{e}_j)\mathbf{w}_i-(\mathbf{e}_j)^T\mathbf{w}_i=0,
\ j \in \{1, 2, 3\},\
i\in\mathcal{N},\label{problem_QCQP.g}\allowdisplaybreaks
\end{align}
where each $\mathbf{e}_{j}$ is a $9\times 1$ standard unit vector
with the $j$th entry being $1$.
By further defining $\mathbf{z}_i\triangleq[\mathbf{w}_i^T 1]^T$,
for $i$ in $\mathcal{N}\cup\{0\}$, and together with the above
matrix form expressions, optimization problem \eqref{problem_new}
can now be transformed into the following equivalent homogeneous
separable QCQP formulation
\begin{align}
\min\limits_{\{\mathbf{z}_i\}}
&\quad \sum_{i=0}^N\mathbf{z}_i^T\mathbf{G}_i\mathbf{z}_i\label{Eq_homo_QCQP}\allowdisplaybreaks\\
\text{s.t.}
&\quad \sum_{k=0}^N\mathbf{z}_k^T\mathbf{G}^c_{ik}\mathbf{z}_k\leq 0,\ i \in\mathcal{N},\label{Eq_homo_QCQP.a}\allowdisplaybreaks\\
&\quad \mathbf{z}_i^T\mathbf{G}^\mu_{i}\mathbf{z}_i\leq 0,\ \mu\in\{u,d,a\},\ i \in\mathcal{N},\label{Eq_homo_QCQP.b}\allowdisplaybreaks\\
%& \ \mathbf{z}^T\mathbf{G}_{ui}\mathbf{z}\leq 0,\ \  i = 1, \ldots, N,\nonumber\\
%& \ \mathbf{z}^T\mathbf{G}_{di}\mathbf{z}\leq 0,\ \  i = 1, \ldots, N,\nonumber\\
%& \ \mathbf{z}^T\mathbf{G}_{ai}\mathbf{z}\leq 0,\ \  i = 1, \ldots, N,\nonumber\\
&\quad \mathbf{z}_i^T\mathbf{G}^P_{i}\mathbf{z}_i=1,\ i \in\mathcal{N},\label{Eq_homo_QCQP.c}\\
&\quad \sum_{i=1}^N\mathbf{z}_i^T\mathbf{G}^{U}_i\mathbf{z}_i\leq C_{\mathrm{UL}},\label{Eq_homo_QCQP.d}\\
&\quad \sum_{i=1}^N\mathbf{z}_i^T\mathbf{G}^{D}_i\mathbf{z}_i\leq C_{\mathrm{DL}},\label{Eq_homo_QCQP.e}\\
&\quad \sum_{i=1}^N\mathbf{z}_i^T\mathbf{G}^{S}_i\mathbf{z}_i\leq C_{\mathrm{Total}},\label{Eq_homo_QCQP.total}\\
&\quad \sum_{i=1}^N\mathbf{z}_i^T\mathbf{G}^{A}_i\mathbf{z}_i\leq f_A,\allowdisplaybreaks\label{Eq_homo_QCQP.f}\\
&\quad \mathbf{z}_i^T\mathbf{G}^{I}_j\mathbf{z}_i=0,\  j \in\{ 1,2,3\},\ i \in\mathcal{N},\label{Eq_homo_QCQP.g}\allowdisplaybreaks\\
&\quad \mathbf{z}_i \succeq 0,\ i \in\mathcal{N}\cup\{0\},
\allowdisplaybreaks\label{Eq_homo_QCQP.h}
\end{align}
where
\begin{align}
&\mathbf{G}_i\triangleq
\begin{bmatrix}
\mathbf{0}& \frac{1}{2}\mathbf{b}_i\\
\frac{1}{2}\mathbf{b}_i^T&0\\
\end{bmatrix},\nonumber\\
&\mathbf{G}^c_{ik}\triangleq
\begin{bmatrix}
\mathbf{0}& \frac{1}{2}\mathbf{b}^c_{ik}\\
\frac{1}{2}{(\mathbf{b}^c_{ik}})^T&0\\
\end{bmatrix},\nonumber\\
&\mathbf{G}^\mu_i\triangleq
\begin{bmatrix}
\mathbf{A}^\mu_{i}&\frac{1}{2}\mathbf{b}^\mu_{i}\\
\frac{1}{2}{(\mathbf{b}^\mu_{i})}^T&0\\
\end{bmatrix},\ \  \mu\in\{u,d,a\},\nonumber\\
&\mathbf{G}^\pi_{i}\triangleq
\begin{bmatrix}
\mathbf{0}&\frac{1}{2}\mathbf{b}^\pi_{i}\\
\frac{1}{2}{(\mathbf{b}^\pi_{i})}^T&0\\
\end{bmatrix},\ \  \pi\in\{P,U,D,S,A\},\nonumber\\
&\mathbf{G}^I_j\triangleq
\begin{bmatrix}
\textrm{diag}(\mathbf{e}_j)&-\frac{1}{2}\mathbf{e}_j\\
-\frac{1}{2}\mathbf{e}_j^T&0\\
\end{bmatrix},\ \ j \in\{ 1,2,3\}.\nonumber
\end{align}

Comparing the optimization problems \eqref{problem_new} and
\eqref{Eq_homo_QCQP}, all constraints have one-to-one corresponding
matrix representations. Specifically, constraint
\eqref{Eq_homo_QCQP.a} is the overall delay constraint, constraint
\eqref{Eq_homo_QCQP.b} comes from the additional auxiliary
constraints \eqref{constraint_D^u}-\eqref{constraint_D^a},
constraint \eqref{Eq_homo_QCQP.c} is the offloading placement
constraint, constraints \eqref{Eq_homo_QCQP.d} and
\eqref{Eq_homo_QCQP.e} correspond to uplink and downlink bandwidth
resource constraints, constraint \eqref{Eq_homo_QCQP.total} is the
total bandwidth constraint, constraint \eqref{Eq_homo_QCQP.f} is the
constraint on the CAP processing resource allocation, and constraint
\eqref{Eq_homo_QCQP.g} corresponds to the integer constraint
\eqref{xy_new}. Therefore, optimization problem \eqref{Eq_homo_QCQP}
is equivalent to the original problem \eqref{e}.

Note that optimization problem \eqref{Eq_homo_QCQP} is a non-convex
separable QCQP problem, which is NP-hard in general. To solve it, we
apply a separable SDR approach to relax it into a separable
semidefinite programming (SDP) problem. Define
$\mathbf{Z}_i\triangleq\zbf_i\zbf_i^T$. We then have
\begin{align}\label{z_to_Z}
\mathbf{z}_i^T\mathbf{G}\mathbf{z}_i =
\textrm{Tr}(\mathbf{G}\mathbf{Z}_i),
\end{align}
with $\textrm{rank}(\mathbf{Z}_i)=1$. By dropping the rank
constraint $\textrm{rank}(\mathbf{Z}_i)=1$, we relax problem
\eqref{Eq_homo_QCQP} into the following separable SDP problem
{\allowdisplaybreaks
\begin{align}
\min\limits_{\{\mathbf{Z}_i\}}
&\quad \sum_{i=0}^N\textrm{Tr}(\mathbf{G}_i\mathbf{Z}_i)\label{Eq_SDP_QCQP}\allowdisplaybreaks\\
\text{s.t.}
&\quad \sum_{k=0}^N\textrm{Tr}(\mathbf{G}^c_{ik}\mathbf{Z}_k)\leq 0,\ i \in\mathcal{N},\label{Eq_SDP_QCQP.a}\allowdisplaybreaks\\
&\quad \textrm{Tr}(\mathbf{G}^\mu_{i}\mathbf{Z}_i)\leq 0,\ \mu\in\{u,d,a\},\ i \in\mathcal{N},\label{Eq_SDP_QCQP.b}\allowdisplaybreaks\\
%& \ \textrm{Tr}(\mathbf{G}_{ui}\mathbf{X})\leq 0,\ \  i = 1, \ldots, N,\nonumber\\
%& \ \textrm{Tr}(\mathbf{G}_{di}\mathbf{X})\leq 0,\ \  i = 1, \ldots, N,\nonumber\\
%& \ \textrm{Tr}(\mathbf{G}_{ai}\mathbf{X})\leq 0,\ \  i = 1, \ldots, N,\nonumber\\
&\quad \textrm{Tr}(\mathbf{G}^P_i\mathbf{Z}_i)=1,\ i \in\mathcal{N},\label{Eq_SDP_QCQP.c}\allowdisplaybreaks\\
&\quad \sum_{i=1}^N\textrm{Tr}(\mathbf{G}^{U}_i\mathbf{Z}_i)\leq C_{\mathrm{UL}},\label{Eq_SDP_QCQP.d}\allowdisplaybreaks\\
&\quad \sum_{i=1}^N\textrm{Tr}(\mathbf{G}^{D}_i\mathbf{Z}_i)\leq C_{\mathrm{DL}},\label{Eq_SDP_QCQP.e}\allowdisplaybreaks\\
&\quad \sum_{i=1}^N\textrm{Tr}(\mathbf{G}^{S}_i\mathbf{Z}_i)\leq C_{\mathrm{Total}},\label{Eq_SDP_QCQP.total}\allowdisplaybreaks\\
&\quad \sum_{i=1}^N\textrm{Tr}(\mathbf{G}^{A}_i\mathbf{Z}_i)\leq f_A,\label{Eq_SDP_QCQP.f}\allowdisplaybreaks\\
%& \ \textrm{Tr}(\mathbf{G}_{D}\mathbf{X})\leq C_{\mathrm{DL}},\nonumber\\
%& \ \textrm{Tr}(\mathbf{G}_{A}\mathbf{X})\leq f_A,\nonumber\\
&\quad \textrm{Tr}(\mathbf{G}^I_j\mathbf{Z}_i)=0,\ j \in\{ 1,2,3\},\ i \in\mathcal{N},\label{Eq_SDP_QCQP.g}\allowdisplaybreaks\\
&\quad \mathbf{Z}_i(10,10)=1,\ i \in\mathcal{N}\cup\{0\},
\label{Eq_SDP_QCQP.h}\\
&\quad \mathbf{Z}_i\succeq 0,\ i \in\mathcal{N}\cup\{0\}.
\label{Eq_SDP_QCQP.i}
\end{align}

The above problem can be solved efficiently in polynomial time using
standard SDP software, such as SeDuMi \cite{grant2009}. Denote
$\mathbf{Z}_i^*$ as the optimal solution of the SDP problem
\eqref{Eq_SDP_QCQP}. We need to obtain the offloading decision
$\mathbf{x}_i$ of the original problem \eqref{e} from
$\mathbf{Z}_i^*$. In the following, we propose a randomization
method to obtain our binary
offloading decisions. %\vspace{-0.2cm}
\subsection{Binary Offloading Decisions via Randomization}\label{sec_randomization}
One might consider using a common approach \cite{luo2010} to obtain
an integer solution from the relaxed SDP problem, by randomly
generating vectors from the Gaussian distribution with zero mean and
covariance $\Zbf_i^*$ for $L$ times, and then mapping them to the
integer set $\{0,1\}^{3N}$ by using the sign of each element in
these vectors. Among the generated vectors, the one that yields the
best objective value of the original problem would be chosen as the
desired solution. However, the above randomization procedure does
not produce a feasible solution due to the offloading decision
placement constraint \eqref{a}. Instead, using the structure of
$\Zbf_i$ and constraints in problem \eqref{Eq_SDP_QCQP}, we propose
the following randomization method for a feasible solution.

 Denote the offloading solution vector as
\begin{align}
\xbf \triangleq [\xbf_1^T,\ldots, \xbf_N^T]^T,\nonumber
\end{align}
where $\xbf_i \triangleq [x^l_{i},x^a_{i},x^c_{i}]^T$, for
$i\in\mathcal{N}$. Since we have removed the rank-1 constraint from
problem \eqref{Eq_homo_QCQP} to arrive at the relaxed problem
\eqref{Eq_SDP_QCQP}, the obtained solution $\Zbf_i^*$ for problem
\eqref{Eq_SDP_QCQP} does not directly provide a feasible binary
solution for the offloading decisions. Our goal is to obtain
appropriate offloading decisions from $\Zbf_i^*$ by mapping its
elements to binary numbers. Note that only the first three elements
in $\zbf_i$ correspond to the offloading decision variables for user
$i$ (see $\wbf_i$ in \eqref{def_w}). Also, since
$\mathbf{Z}_i=\zbf_i\zbf_i^T$ and $\mathbf{z}_i(10)=1$, we know that
the last row of $\mathbf{Z}_i$ satisfies $\Zbf_i(10,j)=\zbf_i(j)$,
for all $j$. Hence, we can use the values of $\Zbf^*_i(10,j)$ to
recover the binary offloading decision $\zbf_i(j)$, for $j=1,2,3$.
Before providing the details of the proposed randomization method,
we first show the property of $\Zbf^*_i(10,j)$, for $j=1,2,3$, in
the following lemma.

\begin{Lem}\label{lem.random}\it
For the optimal solution $\Zbf^*_i$ of problem \eqref{Eq_SDP_QCQP},
$\Zbf^*_i(10,j) \in [0,1]$, for $j=1,2,3$, and $i\in \mathcal{N}$.
%The elements $\Zbf^*_i(10,j)$, for $j=1,2,3$, in $\Zbf^*_i$ are
%always between $0$ and $1$, for $i \in \mathcal{N}$.
%The diagonal term in $\hat{\Zbf}_i^*$ is always between $0$ and $1$,
%for $i \in \mathcal{N}$.
\end{Lem}
\begin{IEEEproof}
After dropping rank-1 constraint $\textrm{rank}(\mathbf{Z}_i)=1$, in
the relaxed problem \eqref{Eq_SDP_QCQP}, constraint
\eqref{Eq_SDP_QCQP.g}
 can only guarantee that $\Zbf_i(j,j)=\Zbf_i(10,j)(=\Zbf_i(j,10))$ for
 $j=1,2,3$. To show that $\Zbf^*_i(10,j)$ $\in [0,1]$, for $j=1,2,3$, we note that, first, $\Zbf_i^*(j,j)\geq 0$
since $\Zbf_i^*\succeq 0$, which means $\Zbf^*_i(10,j)\geq 0$, for
$j=1,2,3$. In addition, constraint \eqref{Eq_SDP_QCQP.c} guarantees
that
\begin{align}
\Zbf^*_i(10,1)+\Zbf^*_i(10,2)+\Zbf^*_i(10,3)=1.\nonumber\allowdisplaybreaks
\end{align}
Thus, we have $\Zbf^*_i(10,j)\in [0,1]$, for $j=1,2,3$.
\end{IEEEproof}

Based on Lemma \ref{lem.random}, we consider a probabilistic mapping
method to find $\mathbf{x}$. We take $\Zbf^*_i(10,j)$ as the
probability of $\zbf_i(j)=1$, i.e.,
$\mathrm{prob}(\zbf_i(j)=1)=\Zbf^*_i(10,j),\ \mathrm{for}\ j=1,2,3$.
Denote
\begin{align}
\mathbf{p}_{i} \triangleq [p^l_{i}, p^a_{i}, p^c_{i}]^T\triangleq
[\Zbf_i^*(10,1),\Zbf_i^*(10,2),\Zbf_i^*(10,3)]^T.\nonumber\allowdisplaybreaks
\end{align}
Equivalently, this means $\mathrm{prob}(x_i^s=1)=p_i^s,\
\mathrm{for}\ s=l,a,c$. We reconstruct $[x_i^l,x_i^a,x_i^c]$ using
$\mathbf{p}_i$ as marginal probabilities,
 while satisfying
constraint \eqref{a}. This leads to our proposed probabilistic
randomization method as follows.

Let
\begin{align}
U^l_i=p^l_i(1-p^a_i)(1-p^c_i),\nonumber\allowdisplaybreaks
\end{align}
\begin{align}
U^a_i=(1-p^l_i)p^a_i(1-p^c_i),\nonumber\allowdisplaybreaks
\end{align}
and
\begin{align}
U^c_i=(1-p^l_i)(1-p^a_i)p^c_i.\nonumber\allowdisplaybreaks
\end{align}
To satisfy the placement constraint \eqref{a}, we define random
vector $\mathbf{u}_i$, which represent the location that user $i$'s
task will be processed, as follows:
\begin{equation}\label{Eq_u_i}
\hspace*{-.15cm}\mathbf{u}_i \hspace*{-.05cm}=\hspace*{-.1cm}
\begin{cases}
\hspace*{-.05cm}[1,0,0]^T\hspace*{-.05cm},&\hspace*{-.25cm}\text{with probability}\ \hspace*{-.05cm}P^l_i\ \hspace*{-.05cm}\text{(local processing)},\\
\hspace*{-.05cm}[0,1,0]^T\hspace*{-.05cm},&\hspace*{-.25cm}\text{with probability}\ \hspace*{-.05cm}P^a_i\ \hspace*{-.05cm}\text{(CAP processing)},\\
\hspace*{-.05cm}[0,0,1]^T\hspace*{-.05cm},&\hspace*{-.25cm}\text{with probability}\ \hspace*{-.05cm}P^c_i\ \hspace*{-.05cm}\text{(cloud processing)},\\
\end{cases}
\end{equation}
where
\begin{align}
P^s_i=\frac{U^s_i}{U^l_i+U^a_i+U^c_i}\ ,\
s\in\{l,a,c\},\nonumber\allowdisplaybreaks
\end{align}
and
\begin{align}
P^l_i+P^a_i+P^c_i=1.\nonumber\allowdisplaybreaks
\end{align}
 We
generate $M$ i.i.d. feasible offloading solutions
$\mathbf{x}^{(m)}=[(\mathbf{u}_1^{(m)})^T\ldots
(\mathbf{u}_N^{(m)})^T]^T$ using the above procedure, for $m = 1,
..., M,$ and solve the corresponding resource allocation problem
\eqref{h} for each $\mathbf{x}^{(m)}$. We then choose among these
feasible solutions the one that gives the minimum objective value of
the optimization problem \eqref{e} to obtain the offloading solution
$\xbf^{\mathrm{sdr}}$ and the corresponding optimal resource
allocation $\{\rbf_i^{\mathrm{sdr}^*}\}$. For the best decision, in
practice, we should also compare $\xbf^{\mathrm{sdr}}$ with the
solutions from local processing only and cloud processing only
methods, and select the one that gives the minimum cost as the final
offloading decision $\xbf^{\mathrm{sdr}^*}$ and the corresponding
optimal resource allocation $\{\rbf_i^{\mathrm{sdr}^*}\}$.

The details of the overall \textit{shareCAP} offloading and resource
allocation algorithm are given in Algorithm \ref{algorithm}. Notice
that the SDP problem \eqref{Eq_SDP_QCQP} can be solved within
precision $\epsilon$ by the interior point method in at most
$O(\sqrt{N}\log(1/\epsilon))$ iterations in which the amount of work
per iteration is $O(N^{6})$ \cite{nesterov1994}. This compares well
with the $3^N$ choices in exhaustive search to find an optimal
offloading decision. In addition, we observe from numerical results
that a small number of randomization trials (e.g., $M=10$) is enough
to give system performance very close to the optimal one.

\begin{Remark} \label{remark processing_cycle}\it The proposed
solution can be easily extended to the general case where the number
of processing cycles for each task on different CPUs are different
because these quantities are constants in optimization problem
\eqref{e}.
\end{Remark}

\begin{algorithm}[!t]
  \caption{\textit{ShareCAP} Offloading Algorithm}
  \begin{algorithmic}[1]
    \State Obtain optimal solution $\Zbf_i^*$ of the SDP problem \eqref{Eq_SDP_QCQP}.
 Extract $\Zbf^*_i(10,j)$, for $j=1,2,3$, from $\mathbf{Z}_i^*$. Set the number of
randomization trials as $M$. \State Record the values of
$\Zbf^*_i(10,j)$, for $j=1,2,3$, as $\pbf_i$.
    \For{$m=1,...,M$}
      \State $\mathbf{x}^{(m)}=[\mathbf{u}_1^{(m)},\ldots, \mathbf{u}_N^{(m)}]^T$ with $\mathbf{u}_i^{(m)}$ generated as in
      \Statex $\ \ \ \ $ \eqref{Eq_u_i};
      \State Given $\mathbf{x}^{(m)}$, solve resource allocation problem \eqref{h}
      \Statex $\ \ \ \ $ and record the minimum cost value of \eqref{h} as $J^{(m)}$;
      \EndFor
    \State Choose among $\xbf^{(1)},\ldots,\xbf^{(M)}$ the one that yields the minimum system cost: $\xbf^{\mathrm{sdr}} = \mathrm{argmin}_{\{\mathbf{x}^{(m)}\}} J^{(m)}$
     \State Compare the minimum cost of \eqref{h} under $\xbf^{\mathrm{sdr}}$ with those under the local processing only and cloud processing only solutions. Select the one that yields the minimum system cost as  $\xbf^{\mathrm{sdr}^*}$.
    \State Output: the proposed offloading solution $\xbf^{\mathrm{sdr}^*}$ and the corresponding optimal resource allocation $\{\rbf_i^{\mathrm{sdr}^*}\}$.
  \end{algorithmic}
  \label{algorithm}
\end{algorithm}

%========================================================================

\section{Offloading with Delay Constraints}\label{sec_delay}
Time-sensitive applications in practice may have strict processing
deadlines, which complicates the offloading decisions and resource
allocation. In this section, we further study the scenario where
each task must be completed before some given deadline. That is,
there is a strict delay constraint for each user's task given by
\begin{equation}\label{eq_delay}
T^L_{i}+T^A_{i}+T^C_{i}\leq T_i,\ i \in\mathcal{N},
\end{equation}
where we note that only one of $T_i^L$, $T_i^A$, and $T_i^C$ is
non-zero by their definitions and constraint \eqref{a}. To ensure
that at least one feasible offloading solution exists, we assume
that local processing time $T^l_i\leq T_i$ so that each user can at
least process its task locally to meet the deadline regardless of
the availability of the remote processing. With above additional
delay constraints, the optimization problem becomes
\begin{align}
\min\limits_{\mathbf{x}, \{ \mathbf{r}_i \} }
&\quad \bigg[\sum_{i=1}^N\rho_i(E^l_{i}x^l_{i}+E^A_ix^a_{i}+E^C_ix^c_{i})\nonumber\\
&\quad +\max_{i}\{T^L_{i}+T^A_{i}+T^C_{i}\}\bigg]\label{problem_delay}\allowdisplaybreaks\\
\text{s.t.} &\quad
\eqref{a},\eqref{b},\eqref{c},\eqref{cap_total},\eqref{d},\eqref{f},\eqref{g},\
\mathrm{and}\ \eqref{eq_delay}.\nonumber
\end{align}
Due to additional delay constraints \eqref{eq_delay}, the
optimization problem \eqref{problem_delay} is more complicated than
the original problem \eqref{e}. In addition, different from
\eqref{e} where any offloading decision is always feasible, only
some offloading decisions are feasible for problem
\eqref{problem_delay}. To solve this problem, in the following, we
modify the original \textit{shareCAP} solution and propose a
three-step algorithm, named \textit{sharedCAP with Delay
Constraints} (\textit{shareCAP-D}). Furthermore, we will show that
the newly obtained binary offloading decision $\mathbf{x}$ and
computation and communication resource allocation $\{ \mathbf{r}_i
\}$ by \textit{shareCAP-D} algorithm are locally optimal.

\subsection{Step 1: QCQP Transformation and Semidefinite Relaxation}
As mentioned above, optimization problem \eqref{problem_delay} is
more complicated, since individual strict delay constraints are
imposed to all users' tasks. Following the similar procedure in
Section \ref{sec_SDR_no_delay}, we move the delay term from the
objective to the constraints by introducing additional auxiliary
variables $t$, and rewrite \eqref{problem_delay} as
\begin{align}\hspace*{-0.3cm}
\min\limits_{\{ \mathbf{x}_i \}, \{ \mathbf{r}_i \},\ t}
&\quad \hspace*{-0.15cm}\sum_{i=1}^N\rho_i(E^l_{i}x^l_{i}+E^A_ix^a_{i}+E^C_ix^c_{i})+t\label{problem_delay_new}\allowdisplaybreaks\\
\text{s.t.}
&\quad \hspace*{-0.15cm}T^l_{i}x^l_{i}+\left(\frac{D_{{\textrm{in}}}(i)}{\eta^u_ic^u_{i}}+\frac{D_{{\textrm{out}}}(i)}{\eta^d_ic^d_{i}}+\frac{Y(i)}{f^a_i}\right)x^a_{i}\nonumber\allowdisplaybreaks\\
&\quad \hspace*{-0.25cm}+\hspace*{-0.05cm}\left(\hspace*{-0.05cm}\frac{D_{{\textrm{in}}}(i)}{\eta^u_ic^u_{i}}\hspace*{-0.05cm}+\hspace*{-0.1cm}\frac{D_{{\textrm{out}}}(i)}{\eta^d_ic^d_{i}}\hspace*{-0.05cm}+\hspace*{-0.05cm}T^{ac}_i\hspace*{-0.1cm}+\hspace*{-0.05cm}T^{c}_i\right)x^c_{i}\hspace*{-0.1cm}\leq \hspace*{-0.05cm}T_i,\  \hspace*{-0.05cm}i\hspace*{-0.05cm}\in\hspace*{-0.05cm}\mathcal{N},\hspace*{-0.05cm}\label{problem_delay_new.a}\allowdisplaybreaks\\
&\quad
\hspace*{-0.15cm}\eqref{a},\eqref{b},\eqref{c},\eqref{cap_total},\eqref{d},\eqref{f},\eqref{xy_new},\
\mathrm{and}\ \eqref{problem_new.a},\nonumber
\end{align}
where constraint \eqref{problem_delay_new.a} comes from the strict
delay constraint \eqref{eq_delay}. Comparing optimization problem
\eqref{problem_delay_new} with problem \eqref{problem_new}, we
observe that they share a similar structure, except that problem
\eqref{problem_delay_new} has the additional delay constraint
\eqref{problem_delay_new.a}. Therefore, we can apply a similar
procedure to transform problem \eqref{problem_delay_new} into a
non-convex separable QCQP problem, and solve the corresponding
separable SDP relaxation problem.

Rewriting the additional constraint \eqref{problem_delay_new.a} into
a matrix form, we have
\begin{align}\label{Eq_homo_QCQP_delay.a}
\mathbf{z}_i^T\mathbf{G}^C_i\mathbf{z}_i\leq T_i,\ i
\in\mathcal{N},
\end{align}
where $\mathbf{z}_i$ is defined as in Section
\ref{sec_SDR_no_delay}, and
\begin{align}
&\mathbf{b}^C_{i}\triangleq[T^l_{i}, 0, (T^{ac}_i+T^c_{i}), 0, 1, 0, 1, 0, 1]^T,\nonumber\allowdisplaybreaks\\
&\mathbf{G}^C_{i}\triangleq
\begin{bmatrix}
\mathbf{0}_{9\times 9}& \frac{1}{2}\mathbf{b}^C_{i}\\
\frac{1}{2}{(\mathbf{b}^C_{i}})^T&0\\
\end{bmatrix}.\nonumber
\end{align}
The optimization problem \eqref{problem_delay_new} can now be transformed
into the following equivalent separable QCQP formulation
\begin{align}%\label{Eq_QCQP}
\min\limits_{\{ \mathbf{z}_i\}}
&\quad \sum_{i=0}^N\mathbf{z}_i^T\mathbf{G}_i\mathbf{z}_i\label{Eq_homo_QCQP_delay}\\
\text{s.t.}
&\quad \eqref{Eq_homo_QCQP.a}-\eqref{Eq_homo_QCQP.h}, \mathrm{and}\ \eqref{Eq_homo_QCQP_delay.a}.\nonumber
\end{align}
Similar to \eqref{z_to_Z}, we have $ \mathbf{Z}_i =
\mathbf{z}_i\mathbf{z}_i^T $ and
\begin{align*}
\mathbf{z}_i^T\mathbf{G}^C_i\mathbf{z}_i=\textrm{Tr}(\mathbf{G}^C_{i}\mathbf{Z}_i).
\end{align*}
Therefore, we can further reformulate problem \eqref{Eq_homo_QCQP_delay} into a
separable SDP problem as follows
\begin{align}
\min\limits_{\{\mathbf{Z}_i\}}
&\quad \sum_{i=0}^N\textrm{Tr}(\mathbf{G}_i\mathbf{Z}_i)\label{Eq_SDP_QCQP_delay}\allowdisplaybreaks\\
\text{s.t.}
&\quad \textrm{Tr}(\mathbf{G}^C_{i}\mathbf{Z}_i)\leq T_i,\ i
\in\mathcal{N},\label{Eq_SDP_QCQP_delay.a}\\
&\quad \eqref{Eq_SDP_QCQP.a}-\eqref{Eq_SDP_QCQP.i},\nonumber
\end{align}
which is SDP problem \eqref{Eq_SDP_QCQP} with the additional delay
constraint \eqref{Eq_SDP_QCQP_delay.a}. Note that SDP problem
\eqref{Eq_SDP_QCQP_delay} is a relaxation of problem
\eqref{problem_delay_new} and always feasible, so that we can obtain
the optimal solution$\{ \mathbf{Z}_i^*\}$. However, the
randomization procedure introduced in Section
\ref{sec_randomization} cannot be directly applied to find a
feasible solution for problem \eqref{problem_delay} due to the
individual delay constraint for each user's task \eqref{eq_delay}.
In other words, there is no feasibility guarantee for the randomly
generated offloading vector $\mathbf{x}$ w.r.t \eqref{eq_delay} and
hence its associated resource allocation problem.

To deal with this issue, we propose a deterministic approach in
which we choose an initial offloading solution that will
subsequently be improved in Steps 2 and 3 below.

\emph{Initial offloading solution:} First, we have $\xbf_i =
[x^l_{i},x^a_{i},x^c_{i}]^T$ and $\mathbf{p}_{i} = [p^l_{i},
p^a_{i}, p^c_{i}]^T$ as defined in Section \ref{sec_randomization}.
Applying Lemma \ref{lem.random}, we can guarantee that $p_i^s \in
[0,1],\ s \in \{ l, a, c\}$. Then, we recover the offloading
decisions $\xbf_i^{\text{sdr}}$ using $\pbf_i$ as follows:
\begin{equation}\label{eq_sdr_recovery}
\hspace*{-.2cm}\mathbf{x}_{i}^{\text{sdr}}
\hspace*{-.05cm}=\hspace*{-.05cm}
\begin{cases}
\hspace*{-.05cm}[1,0,0]^T\hspace*{-.05cm},&\hspace*{-.25cm}\text{if}\   \smash{\displaystyle\max_{s\in \{l,a,c\}}}  p_i^s\hspace*{-.05cm}=p^l_{i}\ \text{(local processing)},\\
\hspace*{-.05cm}[0,1,0]^T\hspace*{-.05cm},&\hspace*{-.25cm}\text{if}\  \smash{\displaystyle\max_{s\in \{l,a,c\}}}  p_i^s\hspace*{-.05cm}=p^a_{i}\ \text{(CAP processing)},\\
\hspace*{-.05cm}[0,0,1]^T\hspace*{-.05cm},&\hspace*{-.25cm}\text{if}\  \smash{\displaystyle\max_{s\in \{l,a,c\}}}  p_i^s\hspace*{-.05cm}=p^c_{i}\ \text{(cloud processing)},\\
\end{cases}
\end{equation}
and obtain the overall offloading decision as
\begin{equation*}
\xbf^{\text{sdr}}=[(\xbf_1^{\text{sdr}})^T,\ldots,
(\xbf_N^{\text{sdr}})^T]^T.
\end{equation*}

Since an offloading decision
$\xbf^{\text{sdr}}$
generated using the above procedure may
not satisfy individual delay constraints \eqref{eq_delay}, in the
following, we introduce an adaptive adjustment procedure to obtain a
feasible solution through iteration, with $\xbf^{\text{sdr}}$ as the
initial solution.

\subsection{Step 2: Obtaining a Feasible Solution via Adaptive Adjustment }

Similar to \eqref{h}, optimization problem \eqref{problem_delay} is
reduced to the optimization of computation and communication
resource allocation $\{ \mathbf{r}_i \}$ given by
\begin{align}
\min\limits_{\{\textbf{r}_i\}}
&\quad \bigg(E+\max_{i}\{T^L_{i}+T^A_{i}+T^C_{i}\}\bigg)\label{resource_problem_delay}\\
\text{s.t.} &\quad
\eqref{b},\eqref{c},\eqref{cap_total},\eqref{d},\eqref{f},
\mathrm{and}\ \eqref{eq_delay},\nonumber
\end{align}
where $E$ is defined below \eqref{h}. We can determine whether a given offloading decision
$\mathbf{x}$ is feasible by solving  problem \eqref{resource_problem_delay} which is convex. If it is feasible, we can obtain
the corresponding optimal resource allocation $\{\mathbf{r}_i\}$.

We now provide an adaptive adjustment procedure to obtain a feasible
offloading solution iteratively. Set
$\xbf^{\text{aa}^*}=\xbf^{\text{sdr}}$. At each iteration:
 \renewcommand{\labelenumi}{\roman{enumi}}
 \begin{enumerate}
 \item Check the
feasibility of $\xbf^{\text{aa}^*}$ by solving problem
\eqref{resource_problem_delay}.
 \item Define a set $N_l^- = \{i:
x_i^l = 0, i\in \mathcal{N}\}$, which contains all users with
current decisions to offload their tasks. If $\xbf^{\text{aa}^*}$ is
infeasible, randomly pick $i \in N_l^-$ and modify the decision to
be local processing as $\mathbf{x}^{\text{aa}^*}_{i}=[1,0,0]^T$.
 \end{enumerate}
Repeat steps i and ii until $\mathbf{x}^{\text{aa}^*}$ is feasible
for problem \eqref{resource_problem_delay}, and record the
corresponding resource allocation as
$\{\mathbf{r}_i^{\text{aa}^*}\}$. Then output the solution of the
adaptive adjustment procedure as
$(\mathbf{x}^{\text{aa}^*},\{\mathbf{r}_i^{\text{aa}^*}\})$.

Note that the above procedure always converges to some feasible
offloading decision $\xbf^{\text{aa}^*}$ (and corresponding optimal
resource allocation $\{ \mathbf{r}_i^{\text{aa}^*} \}$). To see
this, we note that in the worst case, the offloading solutions
$\xbf^{\text{aa}^*}$ converges to the no offloading decision profile
where each task is processed locally, which is
 feasible since $T_i^l \leq T_i$ for $i\in\mathcal{N}$. We summarize this
 property in the following proposition.
 \begin{Prop}\label{lem.AA} \it
$(\xbf^{\text{aa}^*},\{\rbf^{\text{aa}^*}_i\})$ obtained from the
adaptive adjustment procedure is always a feasible solution to the
original optimization problem \eqref{problem_delay} with strict
delay constraints.
\end{Prop}

\subsection{Step 3: Obtaining a Local Optimum via Sequential Turning}
%\vspace{-0.1cm}
With a feasible solution
$(\xbf^{\text{aa}^*},\{\rbf^{\text{aa}^*}_i\})$ obtained in Step 2,
we now propose an iterative procedure, termed \textit{sequential
tuning}, to further reduce the system cost and obtain
 a local optimum for problem \eqref{problem_delay}.

Set
$(\mathbf{x}^{\text{st}^*},\{\rbf^{\text{st}^*}_i\})=(\xbf^{\text{aa}^*},\{\rbf^{\text{aa}^*}_i\})$
as the initial point. At each iteration:
\renewcommand{\labelenumi}{\roman{enumi}}
 \begin{enumerate}
 \item Randomly order the list of all users.
 \item Go through the user list one by one. For each examined user, check
 the three possible offloading decisions for its task, while keep the offloading decisions
 of all other users unchanged. For each offloading decision, find the total system
 cost by solving problem \eqref{resource_problem_delay}. As soon as some user $i$ is found
 to admit a lower total system cost by changing its offloading decision,
 update $(\mathbf{x}^{\text{st}^*},\{\rbf^{\text{st}^*}_i\})$ to the new offloading decision
 and resource allocation that give the lower cost, and exit the
 iteration.
 \end{enumerate}
Repeat steps i and ii until $\mathbf{x}^{\text{st}^*}$ converges,
i.e., no change
 for $\mathbf{x}^{\text{st}^*}$ can be made. Then output the solution of the sequential turning procedure as
$(\mathbf{x}^{\text{st}^*},\{\mathbf{r}_i^{\text{st}^*}\})$.

The above procedure is guaranteed to converge. This is because there
is a finite
 number of possible values for $\mathbf{x}^{\text{st}}_i[t]$. The iteration eventually will reach some
$(\mathbf{x}^{\text{st}^*},\{\mathbf{r}_i^{\text{st}^*}\})$, where
the total system cost cannot be further reduced by modifying any
user's offloading decision (and corresponding resource allocation).
It is straightforward to show that
$(\mathbf{x}^{\text{st}^*},\{\mathbf{r}_i^{\text{st}^*}\})$ is a
local optimum of problem \eqref{problem_delay}, since it gives the
lowest system cost in the joint binary-valued neighborhood of
$\mathbf{x}$ and neighborhood of $\{\mathbf{r}_i\}$. This result is
stated in the following proposition.
%, whose detailed proof is omitted
%due to space limitation.

\begin{Prop}\label{thm.ST} \it
Given any feasible initial point,
$(\mathbf{x}^{\text{st}^*},\{\mathbf{r}_i^{\text{st}^*}\})$ obtained
from the sequential tuning procedure is a local optimal solution to
the original optimization problem \eqref{problem_delay} with strict
delay constraints.
\end{Prop}

We summarize the above three-step \textit{shareCAP-D} algorithm in
Algorithm \ref{algorithm_delay}. Note that, by design, the final
solution $(\xbf^{\text{st}^*},\{\rbf^{\text{st}^*}_i\})$ obtained by
adopting the sequential tuning procedure is better than or at least
as good as $(\xbf^{\text{aa}^*},\{\rbf^{\text{aa}^*}_i\})$. In
Section~\ref{sec_simulation}, we show that the proposed
\textit{shareCAP-D} method provides not only a local optimum
solution but also nearly optimal performance compared with the
optimal policy.

\begin{Remark} \label{remark shareCAP}\it
The sequential tuning procedure can also be applied as an extension
of \textit{shareCAP} for the case without the delay constraints, to
obtain a locally optimal solution with an even lower total system
cost to problem \eqref{e}.
\end{Remark}

\begin{algorithm}[t]
  \caption{\textit{ShareCAP-D} Offloading Algorithm}
  %\color{blue}
  %\small
  \begin{algorithmic}[1]
  \Statex \textbf{Step 1: Initial offloading solution via SDR}
  %\State Transform \eqref{problem_delay} into the SDR
%  problem.
    \State Obtain optimal solution $\{\Zbf_i^*\}$ of the SDR
  problem \eqref{Eq_SDP_QCQP_delay}. Extract the upper-left $3\times 3$ sub-matrices
$\{\hat{\mathbf{Z}}_i^*\}$ from $\{\Zbf_i^*\}$.
 \State Record the values of diagonal terms in
$\hat{\mathbf{Z}}_i^*$ by $\pbf_i=[p^l_{i}, p^a_{i}, p^c_{i}]^T$.
      \State Set $\xbf^{\text{sdr}}=[(\xbf_1^{\text{sdr}})^T,\ldots,
(\xbf_N^{\text{sdr}})^T]^T$, where $\mathbf{x}^{\text{sdr}}_{i}$ is
given by \eqref{eq_sdr_recovery}, as the initial offloading
solution.
    \Statex \textbf{Step 2: Adaptive adjustment}
    \State Set $\mathbf{x}^{\text{aa}^*}=\xbf^{\text{sdr}}$.
    \State Set $\mathrm{AA}=\mathrm{False}$.
    \While{$\mathrm{AA}==\mathrm{False}$}
   \State Check the
feasibility of $\xbf^{\text{aa}^*}$ by solving problem
\eqref{resource_problem_delay};
    \If{$\mathbf{x}^{\text{aa}^*}$ is infeasible}
    \State Determine the set of users with offloaded task:
    \Statex $\ \ \ \ \ \ \ \ \ \ \ $Set $N_l^- = \{i:x_i^l = 0, i\in \mathcal{N}\}$;
    \State Randomly pick user $i \in N_l^-$ and set its offloading
    \Statex $\ \ \ \ \ \ \ \ \ \ \ $decision to local processing: $\mathbf{x}^{\text{aa}^*}_{i}=[1,0,0]^T$;
    \Else
    \State Record the corresponding resource allocation
    \Statex $\ \ \ \ \ \ \ \ \ \ \ $$\{\rbf^{\text{aa}^*}_i\}$;
    \State Set $\mathrm{AA}=\mathrm{True}$;\Comment{Exit while loop}
    \EndIf
    \EndWhile
    \Statex \textbf{Step 3: Sequential tuning}
    \State Set $(\mathbf{x}^{\text{st}^*},\{\rbf^{\text{st}^*}_i\})=(\xbf^{\text{aa}^*},\{\rbf^{\text{aa}^*}_i\})$,
    and record the corresponding total system cost as
$J^{\text{st}^*}$.
    %\State Set $\mathbf{x}^{\text{st}}[0]=\xbf^{\text{aa}^*}$ as
%the initial point for the sequential turning procedure, and
%    solve
%    \eqref{resource_problem_delay} to obtain $\{\mathbf{r}_i^{\text{st}^*}[0]\}$.
    \State Set $\mathrm{ST}=\mathrm{False}$.
\While{$\mathrm{ST}==\mathrm{False}$}
    \State Randomly order the list of all users;
    \State Set user index $j=1$;
    \While{$j\leq N$}
    \State Keep $\mathbf{x}^{\text{st}^*}_{j'}, j'\neq j,\ j'\in \mathcal{N}$ unchanged.
    For the
    \Statex $\ \ \ \ \ \ \ \ \ \ $ three possible offloading choices of ${\mathbf{x}^{\text{st}^*}_j}$,
    find
    \Statex $\ \ \ \ \ \ \ \ \ \ $ their respective total system costs by
    solving
    \Statex $\ \ \ \ \ \ \ \ \ \ $ problem \eqref{resource_problem_delay}; set $J^{\mathrm{st}'}$ as the minimum cost
    \Statex $\ \ \ \ \ \ \ \ \ \ $ among these choices, and record the
    \Statex $\ \ \ \ \ \ \ \ \ \ $ corresponding solution as  $\mathbf{x}^{\mathrm{st}'}_j$ and $\{\mathbf{r}_i^{\mathrm{st}'}\}$;

    %\State Change the offloading decision of user $i$'s task $j$ from
%    \Statex $\ \ \ \ \ \ \ \ $ $\mathbf{x}^{\text{st}^*}_{ij}[t]$ to two other possible choices, and obtain the
%    \Statex $\ \ \ \ \ \ \ \ $ corresponding resource allocations and total
%    system
%    \Statex $\ \ \ \ \ \ \ \ $ costs by solving problem \eqref{h};
%    \State Set the one of these two with the lower total
%    \Statex $\ \ \ \ \ \ \ \ $ system cost as $\mathbf{x}^{\text{st}'}_{ij}$; set the
%    corresponding resource
%    \Statex $\ \ \ \ \ \ \ \ $ allocation and total system cost as $\{\rbf^{\text{st}^*}_i\}$ and $J^{\text{st}'}$;
    \If{$J^{\text{st}'}< J^{\text{st}^*}$}
    \State Set
    $\mathbf{x}^{\text{st}^*}_j=\mathbf{x}^{\text{st}'}_j$, $\{\rbf^{\text{st}^*}_i\}=\{\rbf^{\text{st}'}_i\}$, $J^{\text{st}^*}=J^{\text{st}'}$;
    \State $j \leftarrow N+1$;
    \ElsIf{$j=N$}
    \State $j \leftarrow N+1$; $\mathrm{ST}=\mathrm{True}$;
    \Comment{No change of $\xbf^{\text{st}^*}$ \Statex $\ \ \ \ \ \ \ \ $$\ \ \ \ \ \ \ \ $$\ \ \ \ \ \ \ \ $
    $\ \ \ \ \ \ \ \ $$\ \ \ \ \ \ \ \ $$\ \ \ \ \ \ \ \ $$\ \ \ \ \ \ \ \ $$\ \ \ \ \ \ \ \ \ $can be found; exit}
    \Else
    \State $j \leftarrow j+1$;
    \EndIf
    \EndWhile
    \EndWhile
    \State Output: The offloading decision $\mathbf{x}^{\text{st}^*}$ and the corresponding resource allocation $\{\mathbf{r}_i^{\text{st}^*}\}$.
  \end{algorithmic}
  \label{algorithm_delay}
\end{algorithm}
%\hspace*{-.6cm}
%========================================================================
%\vspace{-0.4cm}

\section{Extension to Sum Delay
Optimization}\label{sec_sum_delay} In previous optimization problems
\eqref{e} and \eqref{problem_delay}, we have considered the maximum
transmission and processing delays among all users as part of the
total system cost. Our approach and proposed solution can also be
extended to the consideration of the sum delay of all users' tasks
as part of the total system cost. This optimization problem is given
as
\begin{align}
\min\limits_{\{ \mathbf{x}_i \}, \{ \mathbf{r}_i \} }
&\quad \sum_{i=1}^N\bigg[\rho_i(E^l_{i}x^l_{i}+E^A_ix^a_{i}+E^C_ix^c_{i})\nonumber\\
&\quad +(T^L_{i}+T^A_{i}+T^C_{i})\bigg]\label{problem_sum_delay}\allowdisplaybreaks\\
\text{s.t.} &\quad
\eqref{a},\eqref{b},\eqref{c},\eqref{cap_total},\eqref{d},\eqref{f},
\mathrm{and}\ \eqref{g}.\nonumber
\end{align}
Same as in problems \eqref{e} and \eqref{problem_delay}, we can
adjust $\rho_i$ to put different emphasis on the sum delay. In
addition, the strictly delay constraint \eqref{eq_delay} can be
included when considering time-sensitive applications.

Using a similar procedure as described in Section
\ref{sec_SDR_no_delay}, we can obtain the corresponding non-convex
separable QCQP problem and separable SDP relaxation problem for
problem \eqref{problem_sum_delay}. The only difference between
problems \eqref{problem_sum_delay} and \eqref{e} is the structure of
the resulting SDR problems. Therefore, the same approach as
\textit{shareCAP} or \textit{shareCAP-D} (when considering the
strictly delay constraint) can again be applied to obtain the final
offloading decision and the corresponding optimal resource
allocation for problem \eqref{problem_sum_delay}. We omit the
derivation details to avoid repetition.

%========================================================================
\section{Numerical Results}\label{sec_simulation}
In this section, we provide numerical results based on Monte Carlo
repeated sampling to study the performance of both \textit{shareCAP}
and \textit{shareCAP-D} under different parameter settings.

\subsection{Parameter Setup}
\begin{figure}[t]
\vspace{-0.1cm} %{\color{blue}
\centering
\includegraphics [scale =0.53]{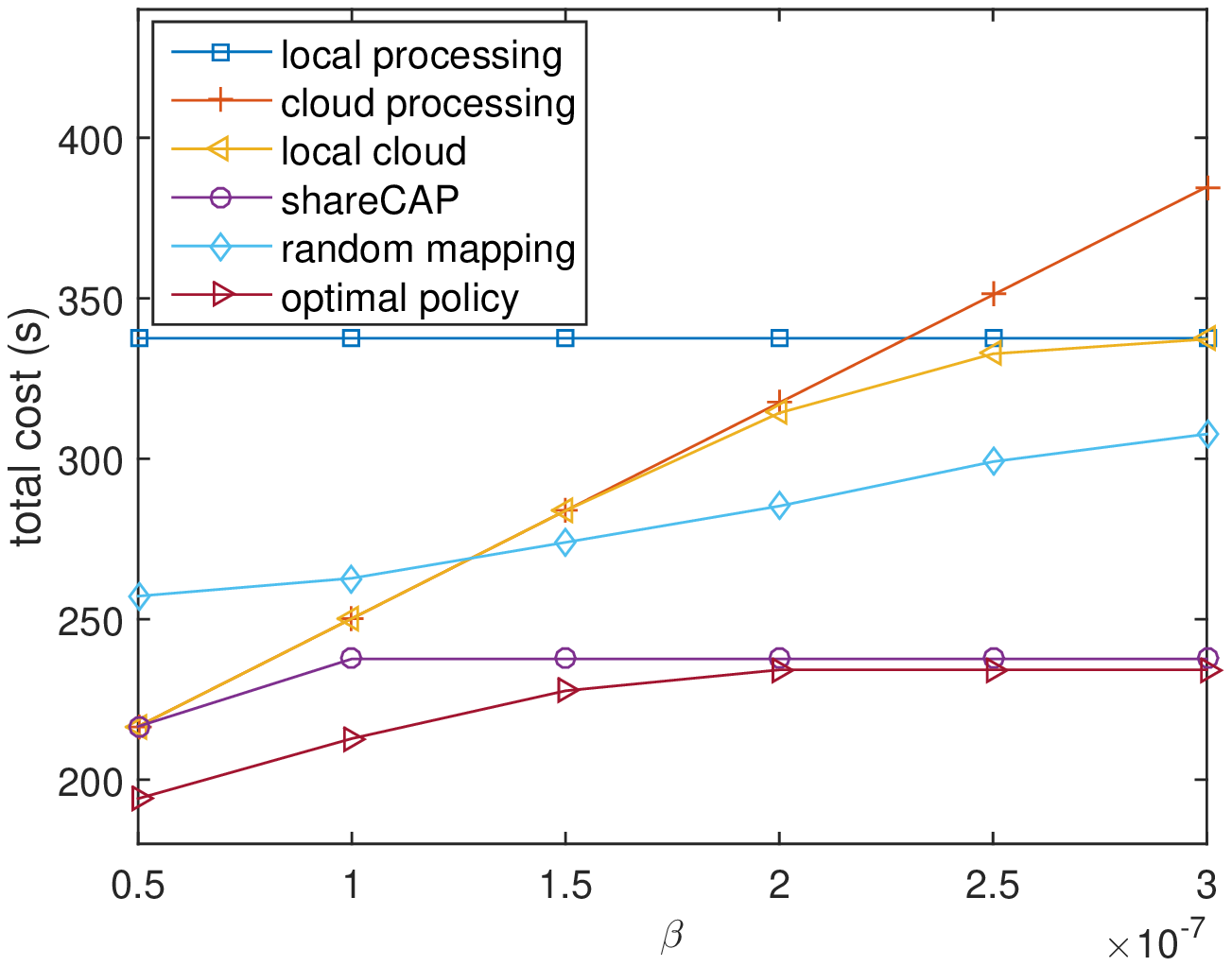}
\caption{The total system cost versus weight $\beta$ (J/bit).
\label{fig:beta}} %\vspace{-0.6cm}
\centering
\includegraphics [scale =0.53]{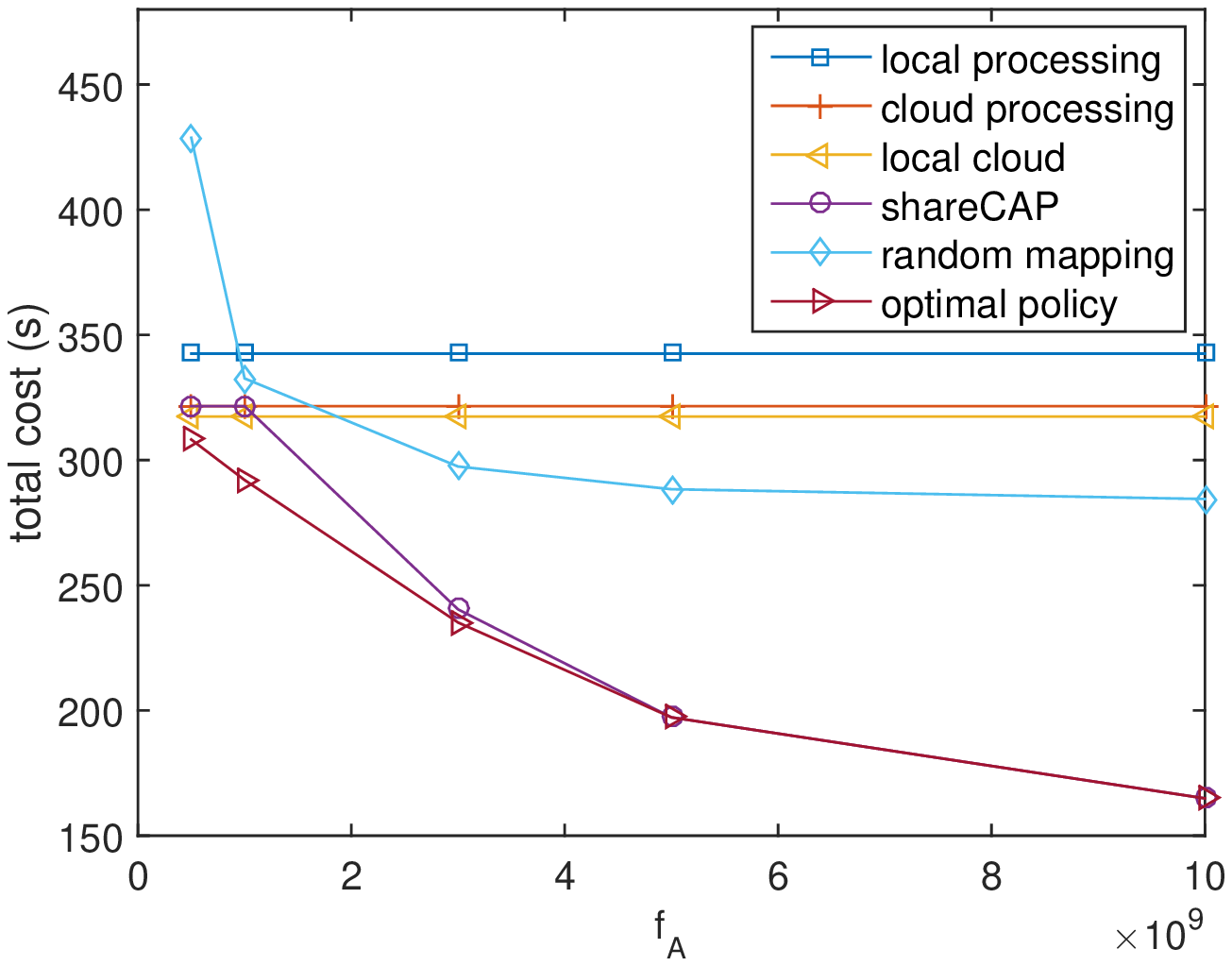}
\caption{The total system cost versus CAP CPU rate $f_A$
(cycles/s).} \label{fig:f_A} \vspace{-0.3cm}
\end{figure}
\begin{figure}[t]
\vspace{-0.1cm} \centering
\includegraphics [scale =0.53]{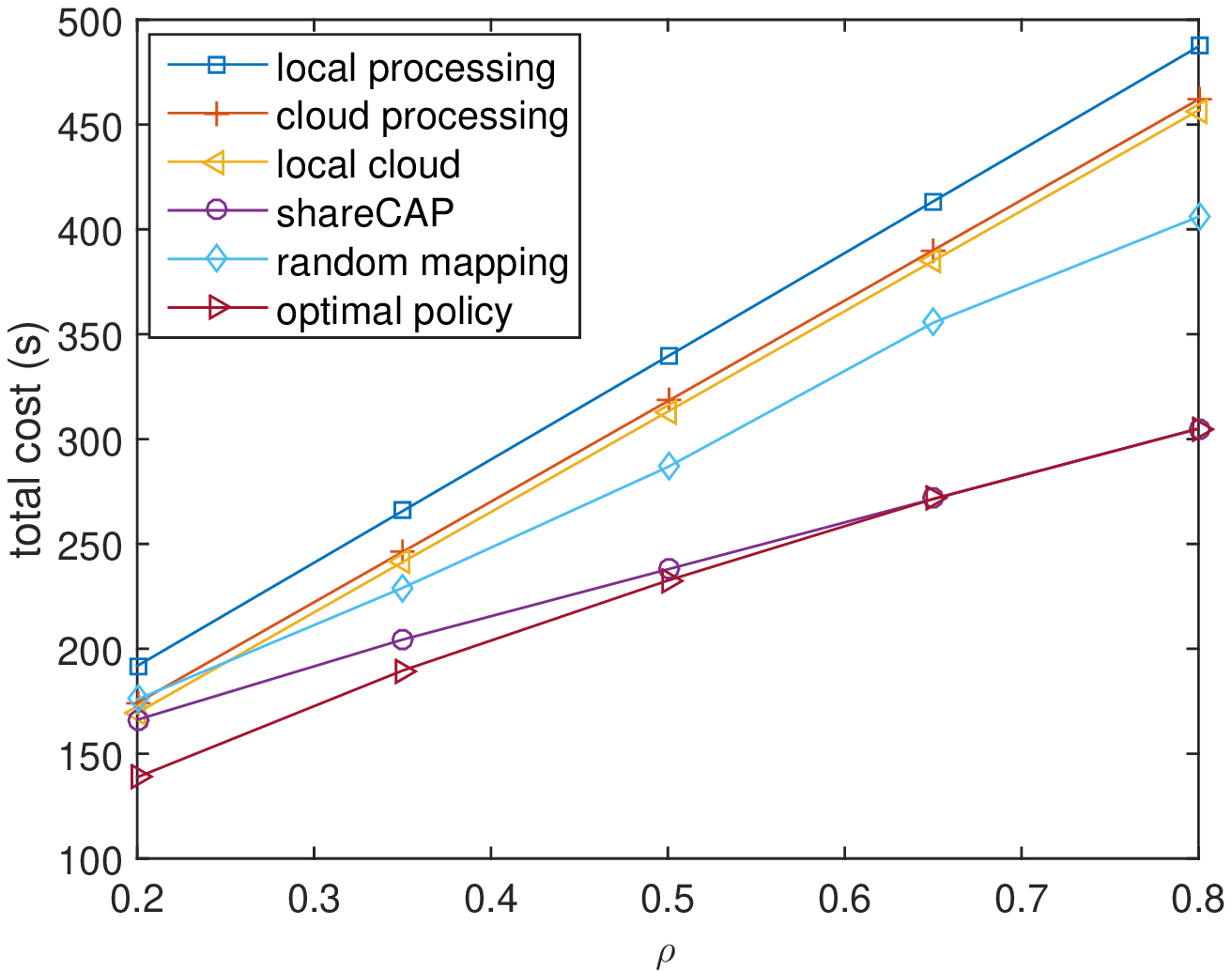}
\caption{The total system cost versus weight $\rho_i=\rho$ (s/J).}
\label{fig:rho}
\centering
\includegraphics [scale =0.53]{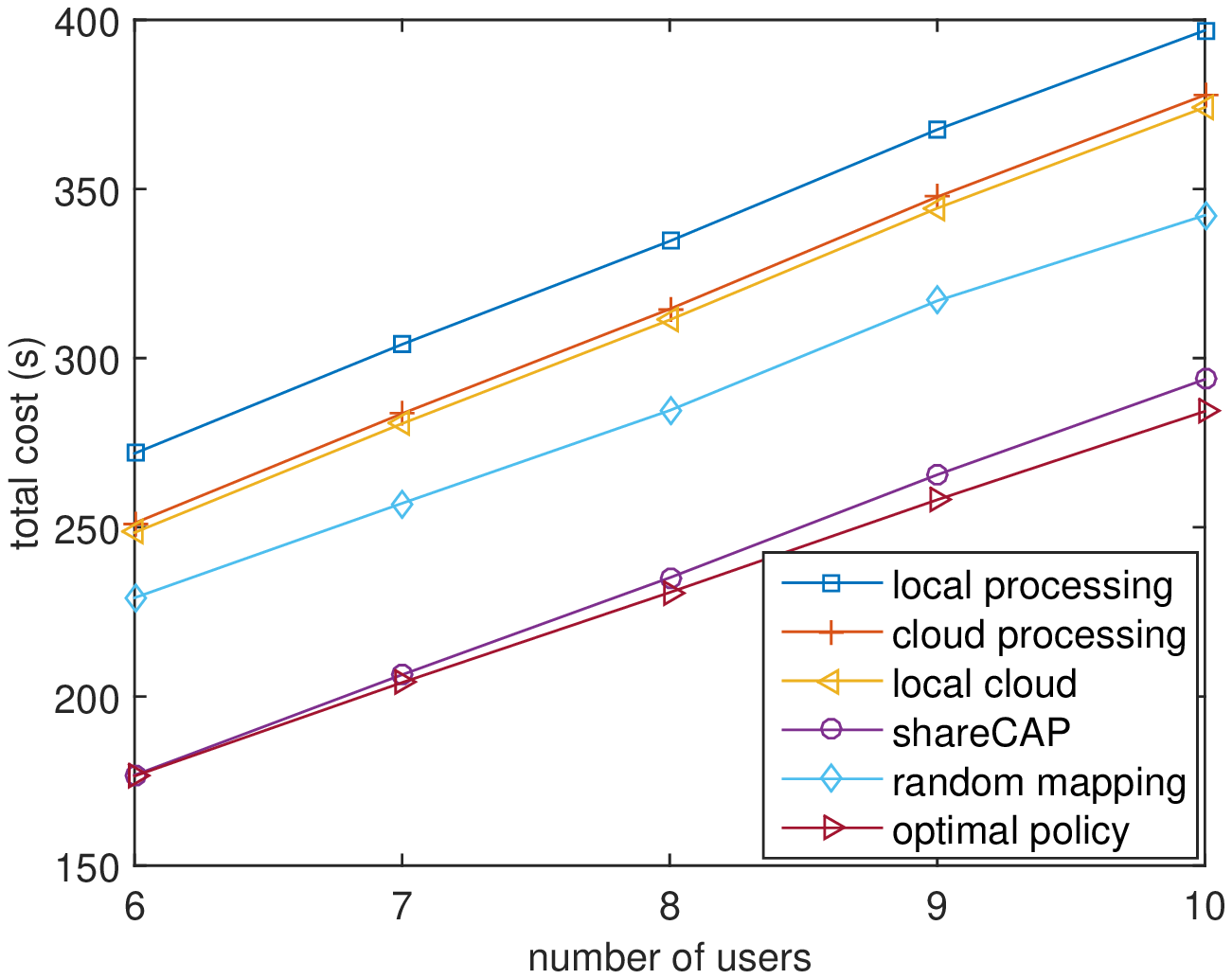}
\caption{The total system cost versus number of users.}
\label{fig:user} \vspace{-0.3cm}
\end{figure}
 In the following, the default
parameter values are described, unless otherwise indicated later. We
adopt the mobile device characteristics from \cite{miettinen2010},
which is based on a Nokia mobile phone, and set the number of users
as $N=8$. According to Tables 1 and 3 in \cite{miettinen2010}, the
mobile device has CPU rate $600\times 10^6$ cycles/s and processing
energy consumption $\frac{1}{650\times 10^6}$ J/cycle, and the local
computation time $3.95\times 10^{-7}$ s/bit and local processing
energy consumption $3.65\times 10^{-7}$ J/bit are calculated when
the x264 CBR encode application (1900 cycles/byte) is considered for
$Y(i)=1900D_{\mathrm{in}}(i)$. The input and output data sizes of
each task are assumed to be uniformly distributed from $10$ to $30$
MB and from $1$ to $3$ MB, respectively.

Both uplink bandwidth $C_{\mathrm{UL}}$ and downlink bandwidth
$C_{\mathrm{DL}}$ between mobile users and the CAP are set to $20$
MHz, with no additional limit on the total bandwidth, and the
transmission and receiving energy consumptions of the mobile user
are both $1.42\times 10^{-7}$ J/bit as indicated in Table 2 in
\cite{miettinen2010}. For simplicity, we set
$\eta^u_{i}=\eta^d_{i}=3.5\ \mathrm{b}/\mathrm{s}/\mathrm{Hz}$ for
all $i$. The CPU rates of the CAP and each server at the remote
cloud are $3\times 10^{9}$ cycle/s and $2\times 10^{9}$ cycle/s,
respectively.
 When tasks are offloaded to the cloud, the
transmission rate $r^{ac}$ is $6$ Mpbs. Also, we set the values of
cost $C^a_{i}$ and $C^c_{i}$ to be the same as that of the input
data size $D_{\textrm{in}}(i)$, and $\alpha = 1\times10^{-8}$ J/bit
and $\beta = 2\times10^{-7}$ J/bit. We further set $\rho_i=\rho=0.5$
s/J for all $i$. Finally, all numerical results are obtained by
averaging over 100 realizations of the input and output data sizes
of each task.

To study the performance of \textit{shareCAP} and
\textit{shareCAP-D}, we compare them with the following methods: 1)
the \textit{local processing only} method where all tasks are
processed by mobile users, 2) the \textit{cloud processing only}
method where all tasks are offloaded to the cloud, 3) the
\textit{local-cloud offloading} method where the same approximation
procedure as the \textit{shareCAP} method is applied except that
there is no CAP, 4) the \textit{random mapping} method where each
task is processed at different locations with equal probability, 5)
the \textit{optimal policy} where the optimal value is obtained by
exhaustive search. When compared with \textit{shareCAP-D}, all of
the adaptive adjustment procedure in Section \ref{sec_delay} is
added to all of the above methods when their offloading decisions
are not feasible.

\begin{figure}[t]
%\vspace{-0.1cm}
\centering
\includegraphics [scale =0.53]{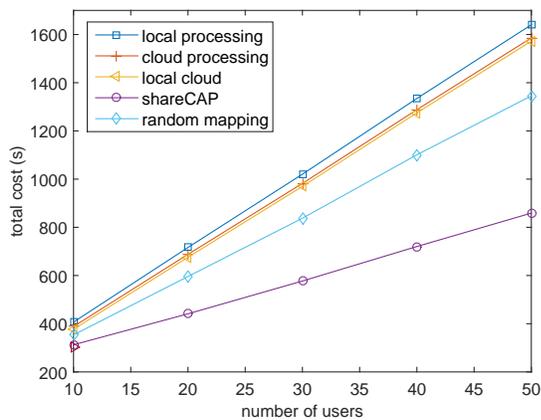}
\caption{The total system cost
versus number of users with scaled resources.} \label{fig:user_scale} %\vspace{+0.1cm}
\end{figure}

\begin{table}[t]
\caption{Average run time comparison under various number of users.}
\centering
\begin{tabular}{|c| c|c|}
\hline %\vspace{0.1cm}
\textbf{Number of} & \textbf{shareCAP (sec)}& \textbf{optimal policy (sec)}\\ \textbf{users} &  &  \textbf{(exhaustive search)}\\
\hline %\vspace{0.2cm}
6 & 1.65  & 146.48 \\\hline
7 & 1.74  & 445.18 \\\hline
8 & 1.90  & 1355.18 \\\hline
9 & 2.07  & 4113.86 \\\hline
10 & 2.22  & 12539.73 \\\hline
20 & 4.14  & N/A\\\hline
30 & 6.56  & N/A\\\hline
40 & 9.82  & N/A\\\hline
50 & 14.37  & N/A\\
 \hline
\end{tabular}\vspace{-0.2cm}
\label{table_run time}
\end{table}

\subsection{Performance of ShareCAP}
In Fig.~\ref{fig:beta}, we show the system cost vs.~weight $\beta$
on the cloud processing cost. When $\beta$ becomes large, the total
system cost puts more emphasis on the cloud usage cost. As a
consequence, all tasks are more likely to be processed by either the
mobile user or the CAP. The \textit{local-cloud} method in this case
converges to the \textit{local processing} method. On the other
hand, when $\beta$ decreases, the cost of \textit{cloud processing}
becomes insignificant, and \textit{shareCAP}, \textit{local-cloud},
and \textit{optimal policy} all converge to \textit{cloud
processing}.

Though the existence of the CAP can provide additional computation
capacity, all tasks processed at the CAP need to share the CAP CPU
rate $f_A$ by optimally allocating the processing rate to each
user's task. In Fig.~\ref{fig:f_A}, we plots the total system cost
vs.~$f_A$. As expected, a more powerful CAP can dramatically
increase system performance, and \textit{shareCAP} converges to
\textit{local-cloud} when the CAP CPU rate is too slow to help.

In Fig.~\ref{fig:rho}, we study the system cost when weight $\rho$
(weight of energy consumption relative to delay) changes. In Figs.
~\ref{fig:user}, and \ref{fig:user_scale}, we study the system cost
under various number of users $N$. In particular, in Fig.
\ref{fig:user_scale}, the amount of limited resources
 (i.e., uplink and downlink bandwidth and the total CAP
processing rate) is scaled proportional to the number of users $N$.
 From Figs.~\ref{fig:rho} to \ref{fig:user_scale}, We observe
that with the help of the CAP, \textit{shareCAP} outperforms all
other methods. Furthermore, all of these figures show that, over a
wide range of system parameter values, \textit{shareCAP} provides
performance close to that of \textit{optimal policy}, where the
latter, obtained by exhaustive search, has an exponential
computational complexity in $N$, i.e., $O(3^N)$. The corresponding
average run times for different values of $N$ are also provided in
Table \ref{table_run time}. They are obtained on a desktop PC with
Intel Core i3-4150 3.5 GHz processor and 8 GB RAM. For
\textit{optimal policy} by exhaustive search, we only obtain the run
times up to 10 users as the required computational time becomes very
high for the cases beyond 10 users.

\begin{figure}[t]
\centering
\includegraphics [scale =0.53]{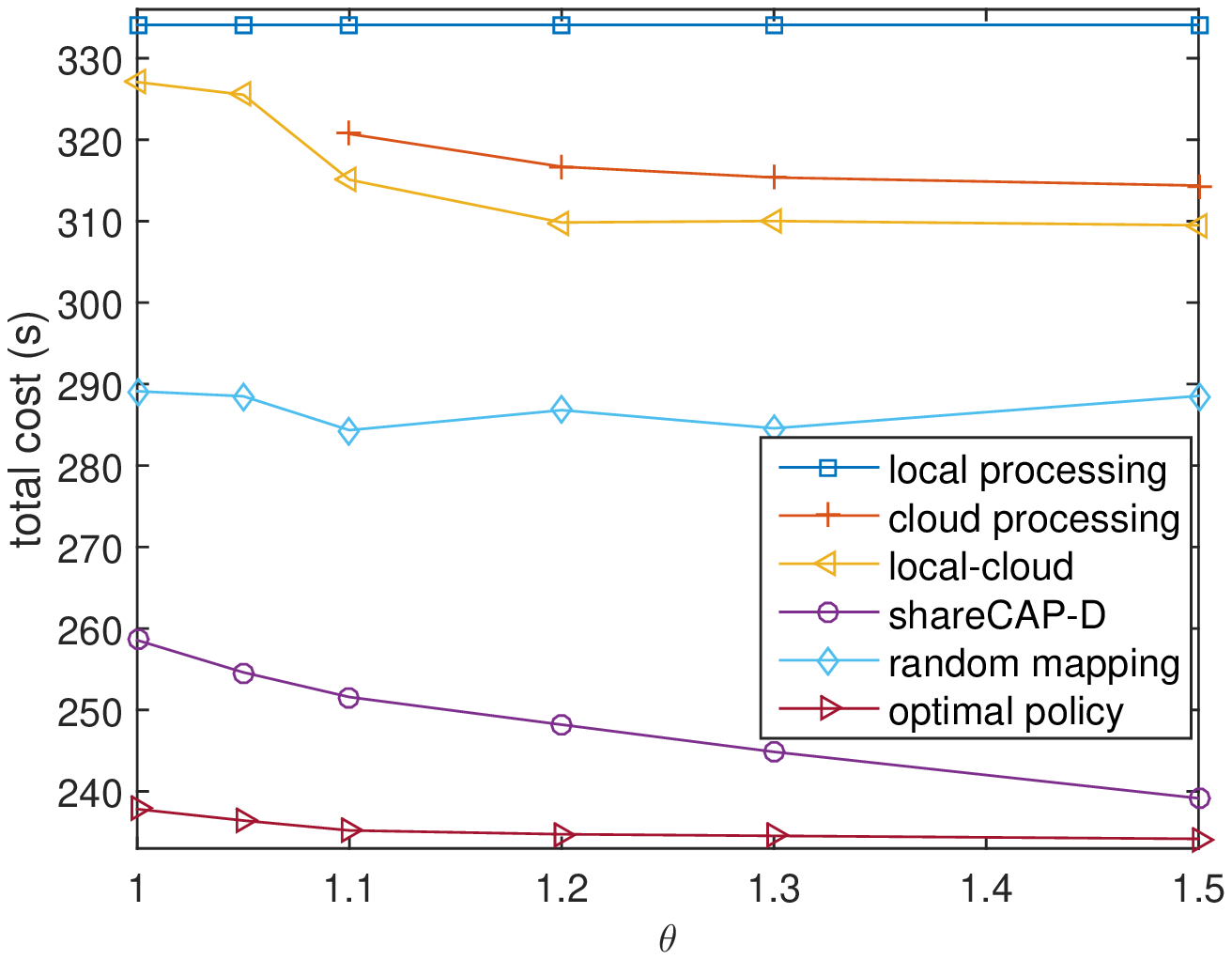}
\caption{The total system cost versus delay factor $\theta$ with strict delay constraints.} \label{fig:theta} %\vspace{-0.5cm}
\centering
\includegraphics [scale =0.53]{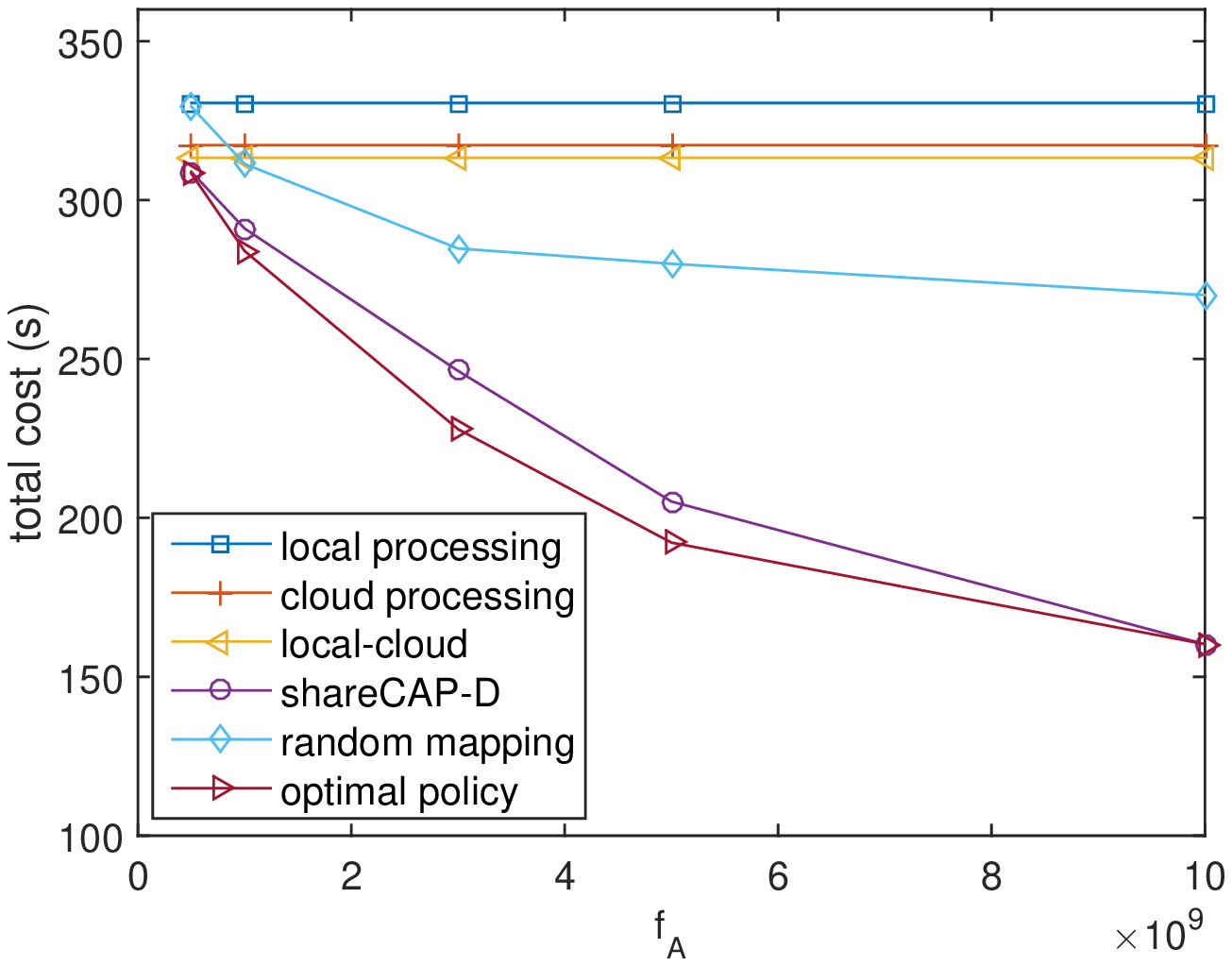}
\caption{The total system cost versus CAP CPU rate $f_A$ (cycles/s)
with strict delay constraints.} \label{fig:fA_delay} \vspace{-0.3cm}
\end{figure}
\subsection{Performance of ShareCAP-D}

For numerical results with \textit{shareCAP-D}, we assume $T_i\geq
\theta T^l_i$ for all $i$.

Fig.~\ref{fig:theta} plots the total system cost vs.~the delay
factor $\theta$. We observe that \textit{shareCAP-D} provides
near-optimal performance unless $\theta$ is close to $1$, i.e., when
the problem \eqref{problem_delay} is nearly infeasible. When
$\theta$ is moderately relaxed, there are more offloading decisions
can satisfy delay constraints, which allows \textit{shareCAP-D} to
choose a local optimum that is closer to the optimal solution.

In Fig.~\ref{fig:fA_delay}, we plots the total system cost vs.~$f_A$
with $\theta=1.1$. We see that \textit{shareCAP-D} outperforms all
other methods except the optimal solution. Furthermore, we observe
that shareCAP-D is the only method that can efficiently utilize the
increasing CAP processing capacity, as demonstrated by it steeply
declining cost curve as $f_A$ increases. In particular, when $f_A
\geq 5\times10^9$, \textit{shareCAP-D} is nearly identical to an
optimal policy.

Finally, we plot the total system cost vs.~number of users $N$ with
$\theta=1.1$ in Fig.~\ref{fig:user_delay}. Though the optimization
problem is more complicated due to additional delay constraints, we
see that the system cost of \textit{shareCAP-D} is still close to
that of the optimal policy, showing the scalability of our proposed
algorithm for various $N$ values.

\begin{figure}[t]
\centering
\includegraphics [scale =0.53]{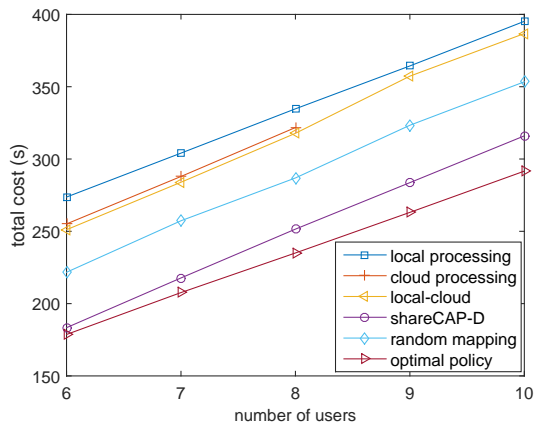}
\caption{The total system cost versus number of users with strict
delay constraints.} \label{fig:user_delay} \vspace{-0.3cm}
\end{figure}

%========================================================================
%\vspace{-0.2cm}
\section{Conclusion}\label{sec_conclusion}
%\vspace{-0.2cm}
We have studied a mobile cloud computing system consisting of
multiple users, one CAP, and one remote cloud server. We propose a
new approach toward joint task offloading and allocation of
computation and communication resources, to minimize the weighted
total cost of energy, computation, and the maximum delay among all
users. Although the optimization problem is non-convex, we propose
\textit{shareCAP}, an efficient heuristic algorithm using SDR and a
new randomization mapping approach. For the case with strict delay
constraints for each task, we propose \textit{shareCAP-D}, a
three-step algorithm to obtain a feasible solution that is locally
optimal. Numerical results suggest that the proposed method gives
nearly optimal performance over a wide range of parameter settings,
and the resultant efficient utilization of a CAP can bring
substantial cost benefit.



\begin{thebibliography}{10}
\providecommand{\url}[1]{#1}
\csname url@samestyle\endcsname
\providecommand{\newblock}{\relax}
\providecommand{\bibinfo}[2]{#2}
\providecommand{\BIBentrySTDinterwordspacing}{\spaceskip=0pt\relax}
\providecommand{\BIBentryALTinterwordstretchfactor}{4}
\providecommand{\BIBentryALTinterwordspacing}{\spaceskip=\fontdimen2\font plus
\BIBentryALTinterwordstretchfactor\fontdimen3\font minus
  \fontdimen4\font\relax}
\providecommand{\BIBforeignlanguage}[2]{{%
\expandafter\ifx\csname l@#1\endcsname\relax
\typeout{** WARNING: IEEEtran.bst: No hyphenation pattern has been}%
\typeout{** loaded for the language `#1'. Using the pattern for}%
\typeout{** the default language instead.}%
\else
\language=\csname l@#1\endcsname
\fi
#2}}
\providecommand{\BIBdecl}{\relax}
\BIBdecl

\bibitem{chen2016icassp}
M.-H. Chen, M.~Dong, and B.~Liang, ``Joint offloading decision and resource
  allocation for mobile cloud with computing access point,'' in \emph{Proc.
  IEEE International Conference on Acoustics, Speech and Signal Processing
  (ICASSP)}, Mar. 2016, pp. 3516--3520.

\bibitem{kumar2013}
K.~Kumar, J.~Liu, Y.-H. Lu, and B.~Bhargava, ``A survey of computation
  offloading for mobile systems,'' \emph{Mobile Networks and Applications},
  vol.~18, no.~1, pp. 129--140, Feb. 2013.

\bibitem{fernando2013}
N.~Fernando, S.~W. Loke, and W.~Rahayu, ``Mobile cloud computing: A survey,''
  \emph{Future Generation Computer Systems}, vol.~29, no.~1, pp. 84 -- 106,
  Jan. 2013.

\bibitem{dinh2013}
H.~T. Dinh, C.~Lee, D.~Niyato, and P.~Wang, ``A survey of mobile cloud
  computing: architecture, applications, and approaches,'' \emph{Wireless
  Communications and Mobile Computing}, vol.~13, no.~18, pp. 1587--1611, 2013.

\bibitem{etsi2016framework}
{ETSI Group Specification}, ``Mobile edge computing ({MEC}); framework and
  reference architecture,'' \emph{ETSI GS MEC 003 V1.1.1}, 2016.

\bibitem{liang2017}
B.~Liang, ``Mobile edge computing,'' in \emph{Key Technologies for 5G Wireless
  Systems}, V. W. S. Wong, R. Schober, D. W. K. Ng, and L.-C. Wang, Eds.,
  Cambridge University Press, 2017.

\bibitem{tran2017}
T.~X. Tran, A.~Hajisami, P.~Pandey, and D.~Pompili, ``Collaborative mobile edge
  computing in 5g networks: New paradigms, scenarios, and challenges,''
  \emph{IEEE Communications Magazine}, vol.~55, no.~4, pp. 54--61, Apr. 2017.

\bibitem{greenberg2008}
A.~Greenberg, J.~Hamilton, D.~A. Maltz, and P.~Patel, ``The cost of a cloud:
  Research problems in data center networks,'' \emph{ACM SIGCOMM Computer
  Communication Review}, vol.~39, no.~1, pp. 68--73, Dec. 2008.

\bibitem{satyanarayanan2009}
M.~Satyanarayanan, P.~Bahl, R.~Caceres, and N.~Davies, ``The case for
  {VM}-based cloudlets in mobile computing,'' \emph{IEEE Pervasive Computing},
  vol.~8, no.~4, pp. 14--23, Oct. 2009.

\bibitem{lewis2015}
G.~Lewis and P.~Lago, ``Architectural tactics for cyber-foraging: Results of a
  systematic literature review,'' \emph{Journal of Systems and Software}, vol.
  107, pp. 158 -- 186, 2015.

\bibitem{bonomi2012}
F.~Bonomi, R.~Milito, J.~Zhu, and S.~Addepalli, ``Fog computing and its role in
  the internet of things,'' in \emph{Proc. ACM SIGCOMM Workshop on Mobile Cloud
  Computing}, Aug. 2012, pp. 13--16.

\bibitem{boyd2004}
S.~Boyd and L.~Vandenberghe, \emph{Convex Optimization}.\hskip 1em plus 0.5em
  minus 0.4em\relax Cambridge University Press, 2004.

\bibitem{luo2010}
Z.-Q. Luo, W.-K. Ma, A.-C. So, Y.~Ye, and S.~Zhang, ``Semidefinite relaxation
  of quadratic optimization problems,'' \emph{IEEE Signal Processing Magazine},
  vol.~27, no.~3, pp. 20--34, May 2010.

\bibitem{kumar2010}
K.~Kumar and Y.-H. Lu, ``Cloud computing for mobile users: Can offloading
  computation save energy?'' \emph{Computer}, vol.~43, no.~4, pp. 51--56, Apr.
  2010.

\bibitem{zhang2013}
W.~Zhang, Y.~Wen, K.~Guan, D.~Kilper, H.~Luo, and D.~Wu, ``Energy-optimal
  mobile cloud computing under stochastic wireless channel,'' \emph{IEEE
  Transactions on Wireless Communications}, vol.~12, no.~9, pp. 4569--4581,
  Sep. 2013.

\bibitem{wen2012}
Y.~Wen, W.~Zhang, and H.~Luo, ``Energy-optimal mobile application execution:
  Taming resource-poor mobile devices with cloud clones,'' in \emph{Proc. IEEE
  International Conference on Computer Communications (INFOCOM)}, Mar. 2012,
  pp. 2716--2720.

\bibitem{barbarossa2013}
S.~Barbarossa, S.~Sardellitti, and P.~Di~Lorenzo, ``Computation offloading for
  mobile cloud computing based on wide cross-layer optimization,'' in
  \emph{Proc. Future Network and Mobile Summit (FutureNetworkSummit)}, Jul.
  2013, pp. 1--10.

\bibitem{munoz2015}
O.~Munoz, A.~Pascual-Iserte, and J.~Vidal, ``Optimization of radio and
  computational resources for energy efficiency in latency-constrained
  application offloading,'' \emph{IEEE Transactions on Vehicular Technology},
  vol.~64, no.~10, pp. 4738--4755, Oct. 2015.

\bibitem{cuervo2010}
E.~Cuervo, A.~Balasubramanian, D.-k. Cho, A.~Wolman, S.~Saroiu, R.~Chandra, and
  P.~Bahl, ``{MAUI}: Making smartphones last longer with code offload,'' in
  \emph{Proc. ACM International Conference on Mobile Systems, Applications, and
  Services (MobiSys)}, Jan. 2010, pp. 49--62.

\bibitem{chun2011}
B.-G. Chun, S.~Ihm, P.~Maniatis, M.~Naik, and A.~Patti, ``Clonecloud: Elastic
  execution between mobile device and cloud,'' in \emph{Proc. ACM Conference on
  Computer Systems (EuroSys)}, Apr. 2011, pp. 301--314.

\bibitem{kosta2012}
S.~Kosta, A.~Aucinas, P.~Hui, R.~Mortier, and X.~Zhang, ``Thinkair: Dynamic
  resource allocation and parallel execution in the cloud for mobile code
  offloading,'' in \emph{Proc. IEEE International Conference on Computer
  Communications (INFOCOM)}, Mar. 2012, pp. 945--953.

\bibitem{zhang2012}
Y.~Zhang, H.~Liu, L.~Jiao, and X.~Fu, ``To offload or not to offload: An
  efficient code partition algorithm for mobile cloud computing,'' in
  \emph{Proc. IEEE International Conference on Cloud Networking (CLOUDNET)},
  Nov. 2012, pp. 80--86.

\bibitem{zhang2013infocom}
W.~Zhang, Y.~Wen, and D.~O. Wu, ``Energy-efficient scheduling policy for
  collaborative execution in mobile cloud computing,'' in \emph{Proc. IEEE
  International Conference on Computer Communications (INFOCOM)}, Apr. 2013,
  pp. 190--194.

\bibitem{mahmoodi2016}
S.~E. Mahmoodi, R.~N. Uma, and K.~P. Subbalakshmi, ``Optimal joint scheduling
  and cloud offloading for mobile applications,'' \emph{IEEE Transactions on
  Cloud Computing}, Apr. 2016.

\bibitem{kao2015}
Y.~H. Kao, B.~Krishnamachari, M.~R. Ra, and F.~Bai, ``Hermes: Latency optimal
  task assignment for resource-constrained mobile computing,'' in \emph{Proc.
  IEEE International Conference on Computer Communications (INFOCOM)}, Apr.
  2015, pp. 1894--1902.

\bibitem{wu2016}
H.~Wu, W.~Knottenbelt, K.~Wolter, and Y.~Sun, ``An optimal offloading
  partitioning algorithm in mobile cloud computing,'' in \emph{Proc.
  International Conference on Quantitative Evaluation of Systems}, Aug. 2016,
  pp. 311--328.

\bibitem{zhang2015}
Y.~Zhang, D.~Niyato, and P.~Wang, ``Offloading in mobile cloudlet systems with
  intermittent connectivity,'' \emph{IEEE Transactions on Mobile Computing},
  vol.~14, no.~12, pp. 2516--2529, Dec. 2015.

\bibitem{Truong-Huu2014}
T.~Truong-Huu, C.~K. Tham, and D.~Niyato, ``To offload or to wait: An
  opportunistic offloading algorithm for parallel tasks in a mobile cloud,'' in
  \emph{Proc. IEEE International Conference on Cloud Computing Technology and
  Science}, Dec. 2014, pp. 182--189.

\bibitem{hoang2012}
D.~T. Hoang, D.~Niyato, and P.~Wang, ``Optimal admission control policy for
  mobile cloud computing hotspot with cloudlet,'' in \emph{Proc. IEEE Wireless
  Communications and Networking Conference (WCNC)}, Apr. 2012, pp. 3145--3149.

\bibitem{Hoang2014}
D.~T. Hoang, D.~Niyato, and L.~B. Le, ``Simulation-based optimization for
  admission control of mobile cloudlets,'' in \emph{Proc. IEEE International
  Conference on Communications (ICC)}, Jun. 2014, pp. 3764--3769.

\bibitem{kaewpuang2013}
R.~Kaewpuang, D.~Niyato, P.~Wang, and E.~Hossain, ``A framework for cooperative
  resource management in mobile cloud computing,'' \emph{IEEE Journal on
  Selected Areas in Communications}, vol.~31, no.~12, pp. 2685--2700, Dec.
  2013.

\bibitem{ren2013}
S.~Ren and M.~van~der Schaar, ``Efficient resource provisioning and rate
  selection for stream mining in a community cloud,'' \emph{IEEE Transactions
  on Multimedia}, vol.~15, no.~4, pp. 723--734, Jun. 2013.

\bibitem{sardellitti2015}
S.~Sardellitti, G.~Scutari, and S.~Barbarossa, ``Joint optimization of radio
  and computational resources for multicell mobile-edge computing,'' \emph{IEEE
  Transactions on Signal and Information Processing over Networks}, vol.~1,
  no.~2, pp. 89--103, Jun. 2015.

\bibitem{lyu2017}
X.~Lyu, H.~Tian, C.~Sengul, and P.~Zhang, ``Multiuser joint task offloading and
  resource optimization in proximate clouds,'' \emph{IEEE Transactions on
  Vehicular Technology}, vol.~66, no.~4, pp. 3435--3447, Apr. 2017.

\bibitem{chen2014}
X.~Chen, ``Decentralized computation offloading game for mobile cloud
  computing,'' \emph{IEEE Transactions on Parallel and Distributed Systems},
  vol.~26, no.~4, pp. 974--983, Apr. 2015.

\bibitem{meskar2017}
E.~Meskar, T.~D. Todd, D.~Zhao, and G.~Karakostas, ``Energy aware offloading
  for competing users on a shared communication channel,'' \emph{IEEE
  Transactions on Mobile Computing}, vol.~16, no.~1, pp. 87--96, Jan. 2017.

\bibitem{chen2015efficient}
X.~Chen, L.~Jiao, W.~Li, and X.~Fu, ``Efficient multi-user computation
  offloading for mobile-edge cloud computing,'' \emph{IEEE/ACM Transactions on
  Networking}, vol.~24, no.~5, pp. 2795--2808, Oct 2016.

\bibitem{chen2016icc}
M.-H. Chen, B.~Liang, and M.~Dong, ``Joint offloading decision and resource
  allocation for multi-user multi-task mobile cloud,'' in \emph{Proc. IEEE
  International Conference on Communications (ICC)}, May 2016.

\bibitem{viswanathan2016}
H.~Viswanathan, P.~Pandey, and D.~Pompili, ``Maestro: Orchestrating concurrent
  application workflows in mobile device clouds,'' in \emph{Proc. IEEE
  International Conference on Autonomic Computing (ICAC)}, Jul. 2016, pp.
  257--262.

\bibitem{Habak2015}
K.~Habak, M.~Ammar, K.~A. Harras, and E.~Zegura, ``Femto clouds: Leveraging
  mobile devices to provide cloud service at the edge,'' in \emph{Proc. IEEE
  International Conference on Cloud Computing}, Jun. 2015, pp. 9--16.

\bibitem{Rahimi2012}
M.~R. Rahimi, N.~Venkatasubramanian, S.~Mehrotra, and A.~V. Vasilakos,
  ``Mapcloud: Mobile applications on an elastic and scalable 2-tier cloud
  architecture,'' in \emph{Proc. IEEE/ACM Fifth International Conference on
  Utility and Cloud Computing}, Nov. 2012, pp. 83--90.

\bibitem{Rahimi2013}
M.~R. Rahimi, N.~Venkatasubramanian, and A.~V. Vasilakos, ``Music:
  Mobility-aware optimal service allocation in mobile cloud computing,'' in
  \emph{Proc. IEEE International Conference on Cloud Computing}, Jun. 2013, pp.
  75--82.

\bibitem{Song2014}
J.~Song, Y.~Cui, M.~Li, J.~Qiu, and R.~Buyya, ``Energy-traffic tradeoff
  cooperative offloading for mobile cloud computing,'' in \emph{Proc. IEEE
  International Symposium of Quality of Service (IWQoS)}, May 2014, pp.
  284--289.

\bibitem{cardellini2016}
V.~Cardellini, V.~De~Nitto~Person{\'e}, V.~Di~Valerio, F.~Facchinei, V.~Grassi,
  F.~Lo~Presti, and V.~Piccialli, ``A game-theoretic approach to computation
  offloading in mobile cloud computing,'' \emph{Mathematical Programming}, vol.
  157, no.~2, pp. 421--449, 2016.

\bibitem{chen2017infocom}
M.-H. Chen, B.~Liang, and M.~Dong, ``Joint offloading and resource allocation
  for computation and communication in mobile cloud with computing access
  point,'' in \emph{Proc. IEEE International Conference on Computer
  Communications (INFOCOM)}, May 2017.

\bibitem{Shmoys1995}
D.~B. Shmoys, J.~Wein, and D.~P. Williamson, ``Scheduling parallel machines
  on-line,'' \emph{SIAM J. Comput.}, vol.~24, no.~6, pp. 1313--1331, Dec. 1995.

\bibitem{grant2009}
\BIBentryALTinterwordspacing
M.~Grant, S.~Boyd, and Y.~Ye, ``{CVX}: {M}atlab software for disciplined convex
  programming,'' 2009. [Online]. Available: \url{http://cvxr.com/cvx/}
\BIBentrySTDinterwordspacing

\bibitem{nesterov1994}
Y.~Nesterov, A.~Nemirovskii, and Y.~Ye, \emph{Interior-point polynomial
  algorithms in convex programming}.\hskip 1em plus 0.5em minus 0.4em\relax
  SIAM, 1994.

\bibitem{miettinen2010}
A.~P. Miettinen and J.~K. Nurminen, ``Energy efficiency of mobile clients in
  cloud computing,'' in \emph{Proc. USENIX Conference on Hot Topics in Cloud
  Computing (HotCloud)}, Jun. 2010, pp. 4--11.

\end{thebibliography}
\end{document}